\documentclass[onecolumn,usenatbib]{mnras}

\usepackage{newtxtext,newtxmath}
\usepackage[T1]{fontenc}
\usepackage{ae,aecompl}
\usepackage{graphicx}	
\usepackage{amsmath}	
\usepackage{amssymb}	

\newcommand{\rd}{\mathrm{d}}

\newcommand{\re}{\mathrm{e}}
\newcommand{\ri}{\mathrm{i}}
\newcommand{\bftheta}{{\mbox{\boldmath $\theta$}}}
\newcommand{\bfmu}{{\mbox{\boldmath $\mu$}}}
\newcommand{\bfJ}{\mathbfit{J}}
\newcommand{\bfL}{\mathbfit{L}}
\newcommand{\bfr}{\mathbfit{r}}
\newcommand{\bfe}{\mathbfit{e}}
\newcommand{\bfs}{\mathbfit{s}}
\newcommand{\bfw}{\mathbfit{w}}
\newcommand{\bfx}{\mathbfit{x}}
\newcommand{\bfy}{\mathbfit{y}}
\newcommand{\bfz}{\mathbfit{z}}

\newcommand{\p}{\upartial}
\newcommand{\egr}{\epsilon_\mathrm{GR}}
\newcommand{\msun}{\,\mathrm{M}_{\sun}}
\newcommand{\pc}{\,\mbox{pc}}
\newcommand{\yr}{\,\mbox{yr}}
\newcommand{\half}{\textstyle{\frac{1}{2}}}

\title[Phase transition in black-hole star clusters]{Order-disorder phase
  transition in black-hole star clusters -- III. A mono-energetic cluster}

\author[]{Scott Tremaine\thanks{tremaine@ias.edu}\\
 Institute for Advanced Study, Princeton, NJ 08540, USA\\}

\begin{document}
\label{firstpage}
\pagerange{\pageref{firstpage}--\pageref{lastpage}}
\maketitle

\begin{abstract}
  Supermassive black holes at the centres of galaxies are often
  surrounded by dense star clusters. For a wide range of cluster
  properties and orbital radii the resonant relaxation times in these
  clusters are much shorter than the Hubble time. Since resonant
  relaxation conserves semimajor axes, these clusters should be in the
  maximum-entropy state consistent with the given semimajor axis
  distribution. We determine these maximum-entropy equilibria in a
  simplified model in which all of the stars have the same semimajor
  axes. We find that the cluster exhibits a phase transition from a
  disordered, spherical, high-temperature equilibrium to an ordered
  low-temperature equilibrium in which the stellar orbits have a
  preferred orientation or line of apsides. Here `temperature' is a
  measure of the non-Keplerian or self-gravitational energy of the
  cluster; in the spherical state, temperature is a function of the
  rms eccentricity of the stars. We explore a simple two-parameter
  model of black-hole star clusters -- the two parameters are
  semimajor axis and black-hole mass --- and find that clusters are
  susceptible to the lopsided phase transition over a range of
  $\sim 10^2$ in semimajor axis, mostly for black-hole masses
  $\lesssim 10^{7.5}\msun$.
\end{abstract}


\section{Introduction}

\noindent
This is one of a series of papers investigating the thermodynamic
equilibria of a black-hole star cluster, by which we mean a stellar
system of mass $M_\star$ orbiting a central black hole of mass
$M_\bullet\gg M_\star$ \citep{ttk19,tre19}. We investigate the
equilibria of these systems on time-scales that are much longer than
the resonant-relaxation time but shorter than the two-body relaxation
time. On these time-scales the semimajor axes of the stars are frozen,
but the eccentricities and orbit orientations are distributed in a
maximum-entropy state in the phase space at a given semimajor
axis. For simplicity we shall make a number of assumptions and
simplifications, of which the most important is that we focus on a
cluster composed of stars at a single semimajor axis. We call this a
`mono-energetic' cluster since the Keplerian energy\footnote{We do not
  distinguish `energy' and `energy per unit mass' in this paper; in
  other words we often assume that the stellar mass is unity. The
  meaning should be clear from the context or dimensional analysis.}
is the same for all stars. We also assume that all stars have the same
mass and ignore the destruction of stars by the black hole.

The assumption of a mono-energetic cluster is unrealistic, but such
clusters provide a fairly simple limiting case that illuminates the
fairly complex dynamics we shall encounter. Moreover, the
mono-energetic cluster has a radial distribution of stars that is very
different from the scale-free cluster investigated by \cite{tre19}, so
we may expect that the behaviour common to these two over-simplified
model systems is also found in clusters with more realistic radial
profiles.  Our focus on mono-energetic clusters was stimulated and
informed by numerical simulations with mono-energtic clusters of
wires, which were reported briefly in \cite{ttk19} and which will be
the subject of a forthcoming paper (Touma \& Kazandjian, in preparation). 

Sections \ref{sec:eq} and \ref{sec:num} describe the analytic and
numerical machinery we use to construct and describe maximum-entropy
stellar systems. The properties of the equilibria, with and without
corrections for relativistic precession, are derived in
\S\ref{sec:results}. Section \ref{sec:disc} sets these results in
context using a simple approximate model of actual black-hole star
clusters. The paper is summarized in \S\ref{sec:summary}. The Appendix
contains calculations of the linear stability of spherical equilibria,
both thermodynamic and dynamical. Some of the results of this paper
have been summarized previously in \cite{ttk19}.

\section{Equilibria of maximum-entropy systems}

\label{sec:eq}

\subsection{Phase-space variables}

\noindent
Let the mass of the central object be $M_\bullet$. The usual Keplerian
orbital elements include semimajor axis $a$, eccentricity $e$,
inclination $I$, argument of periapsis $\omega$, and angle of the
ascending node $\Omega$. The angular momentum per unit mass
$L\equiv (GM_\bullet a)^{1/2}(1-e^2)^{1/2}$ and the $z$-component of the angular
momentum $L_z\equiv L\cos I$. The position of a particle in its
orbit can be specified by the mean anomaly $\ell$, the eccentric
anomaly $u$, or the true anomaly $f$.

We define a set of action-angle variables by the actions
$\Lambda\equiv (GM_\bullet a)^{1/2}$, $L$, and $L_z$ and the conjugate
angles $\ell$, $\omega$, and $\Omega$. Let
$(\bfL,\bftheta)=(L,L_z,\omega,\Omega)$; these variables vary during
resonant relaxation while $\Lambda$ is constant.

The canonical volume element in phase space is
\begin{align}
\rd\bfmu &= \rd\Lambda \rd L \rd L_z \rd\omega \rd\Omega \rd\ell\equiv
     \rd\Lambda \rd\bfL\rd\bftheta \rd\ell\nonumber \\
&=\textstyle{\frac{1}{4}}(GM_\bullet)^{3/2} a^{1/2} \rd a \rd e^2\sin I \rd I \rd\omega
\rd\Omega \rd\ell.
\label{eq:mudef}
\end{align}

The equilibrium mass distribution function (hereafter DF) in a Kepler
potential may be written $F(\Lambda,\bfL,\bftheta)$, since Jeans's
theorem ensures that it is independent of the mean anomaly $\ell$. In
this paper, the DF is normalized such that
$F(\Lambda,\bfL,\bftheta)\rd\bfmu$ is the mass in the phase-space
volume element $\rd\bfmu$.

\subsection{General relativity}

\label{sec:gr}

\noindent
The most important effect of general relativity on stellar orbits is apsidal precession, which takes place at an
orbit-averaged rate
\begin{equation}
  \dot\omega_\mathrm{GR}=\frac{3(GM_\bullet)^{3/2}}{c^2a^{5/2}(1-e^2)}=\frac{3(GM_\bullet)^{5/2}}{c^2a^{3/2}L^2}.
\label{eq:omegadot}
\end{equation}
The corresponding Hamiltonian is found by integrating $\dot\omega_\mathrm{GR}=\p
H_\mathrm{GR}/\p L$, which yields 
\begin{equation}
  H_\mathrm{GR}(\Lambda,L)=-\frac{3(GM_\bullet)^{5/2}}{c^2a^{3/2}L}=-\frac{3(GM_\bullet)^4}{c^2\Lambda^3L}
\label{eq:hgr}
\end{equation}
plus a term that is unimportant (for our purposes) depending on $a$ but not
$L$. 

Equation (\ref{eq:omegadot}) shows that the apsidal precession due to
general relativity $\dot\omega \propto L^{-2}$, which diverges for radial
orbits. This divergence has two important and related consequences:

\begin{enumerate}

\item The rapid precession suppresses resonant relaxation, which
  therefore becomes less effective than two-body relaxation
  for nearly radial orbits (the `Schwarzschild barrier'; see
  \citealt{mer11,boa16,alex17}). Thus the fundamental approximation
  on which our analysis is based -- the existence of a metastable
  equilibrium on time-scales long compared to the resonant relaxation
  time but short compared to the two-body relaxation time -- fails for
  high-eccentricity orbits.

\item In a canonical ensemble with inverse temperature $\beta$, the DF
  is proportional to $\exp(-\beta H)$ where $H$ is the
  Hamiltonian. Since $H_\mathrm{GR}$ is negative and divergent as $L\to
  0$, the DF also diverges if the temperature is positive. 

\end{enumerate}

The nature of the relaxation process near the Schwarzschild barrier
has been described in detail for spherical systems in the references
above. We shall not attempt a similar treatment for non-spherical
systems. Instead, when relativistic effects are included we simply
truncate the phase space at some maximum eccentricity $e_{\max}$ that
is intended to represent the upper limit to the phase-space region in
which resonant relaxation is more effective than two-body relaxation.

\subsection{Entropy and energy}

\noindent
The entropy is 
\begin{equation} 
S=- 2\pi \int \rd\Lambda \rd\bfL \rd\bftheta\,
F(\Lambda,\bfL,\bftheta) \log F(\Lambda,\bfL,\bftheta).
\label{eq:entropy}
\end{equation} 

Since semimajor axes are conserved in resonant relaxation, the
Keplerian energy $E=-\frac{1}{2} GM_\bullet m/a$ of each star is
conserved. Therefore for brevity we use the term `energy' to denote
the non-Keplerian component of the total energy, which arises from the
relativistic Hamiltonian (\ref{eq:hgr}) and from the orbit-averaged
gravitational interactions between the stars:
\begin{align}
E\equiv 2\pi\int \rd\Lambda \rd\bfL \rd\bftheta\, F(\Lambda,\bfL,\bftheta)
H_\mathrm{GR}(\Lambda,L) + 2\pi^2\int \rd\Lambda \rd\bfL \rd\bftheta \rd\Lambda' \rd\bfL'
\rd\bftheta'\, F(\Lambda,\bfL,\bftheta)
\mathbfss{K}(\Lambda,\Lambda',\bfL,\bfL',\bftheta,\bftheta') F(\Lambda',\bfL',\bftheta'),
\label{eq:edef}
\end{align}
where $\mathbfss{K}$ is the time-averaged gravitational potential energy between
unit masses on distinct Keplerian orbits, 
\begin{equation}
\mathbfss{K}(\Lambda,\Lambda',\bfL,\bfL',\bftheta,\bftheta')=-G\int
\frac{\rd\ell}{2\pi}\frac{\rd\ell'}{2\pi}\,\frac{1}{|\bfr-\bfr'|}.
\label{eq:kkdef}
\end{equation}

The Hamiltonian of the system is\footnote{Note that in our convention
  the Hamiltonian has units $(\mbox{velocity})^2$.}
\begin{equation}
H(\Lambda,\bfL,\bftheta)=H_\mathrm{GR}(\Lambda,L) + 2\pi\int \rd\Lambda' \rd\bfL'
\rd\bftheta'\,\mathbfss{K}(\Lambda,\Lambda',\bfL,\bfL',\bftheta,\bftheta') F(\Lambda',\bfL',\bftheta').
\label{eq:hdef}
\end{equation}
The entropy at fixed energy and mass is extremized when the DF has the
form 
\begin{equation}
F(\Lambda,\bfL,\bftheta)=A(\Lambda)\exp[-\beta
H(\Lambda,\bfL,\bftheta)]
\label{eq:betadef}
\end{equation}
where $\beta$ is an inverse temperature. Since resonant relaxation
does not affect semimajor axes, the mass per unit semimajor axis
\begin{equation}
\rho(\Lambda)\equiv 2\pi\int \rd\bfL \rd\bftheta\,F(\Lambda,\bfL,\bftheta)
\label{eq:rhodef}
\end{equation}
is conserved. Thus the function $A(\Lambda)$ in a maximum-entropy
state is determined by the non-linear equation 
\begin{equation}
A(\Lambda)=\frac{\rho(\Lambda)}{2\pi\int  \rd\bfL \rd\bftheta\, \exp[-\beta
H(\Lambda,\bfL,\bftheta)]},
\end{equation}
in which $H(\Lambda,\bfL,\bftheta)$ depends on $A(\Lambda)$ through
equation (\ref{eq:hdef}). 

\subsection{The averaged gravitational potential} 

\noindent
To evaluate the kernel $\mathbfss{K}$ we use the standard expansion in
spherical coordinates $\bfr=(r,\theta,\phi)$
\begin{equation}
\frac{1}{|\bfr-\bfr'|}=\sum_{l=0}^\infty\sum_{m=-l}^l
\frac{4\pi}{2l+1} \frac{r_<^l}{r_>^{l+1}}Y_{lm}^\ast(\theta',\phi')Y_{lm}(\theta,\phi),
\label{eq:std}
\end{equation}
where $r_<$ and $r_>$ are the smaller and larger of $r$ and $r'$ and
$Y_{lm}(\theta,\phi)$ is a spherical harmonic.  We also use the
representation of a spherical harmonic in orbital elements 
\begin{equation}
 Y_{lm}(\theta,\phi)=\sum_{n=-l}^l \ri^{n-m}d^{\,l}_{nm}(I)Y_{ln}(\half\pi,0)\re^{\ri m\Omega}\re^{\ri n(f+\omega)}.
\label{eq:ylm}
\end{equation}
Here $f$ is the true anomaly and the Wigner d-matrix is 
\begin{equation}
d^{\,l}_{nm}(I)\!=\!\sum_s\!\frac{(-1)^s\sqrt{(l+n)!(l-n)!(l+m)!(l-m)!}}{(l+m-s)!(l-n-s)!s!(s+n-m)!}\!\left(\cos\half
I\right)^{2l+m-n-2s}\left(\sin\half I\right)^{2s+n-m}\!\!,
\label{eq:wigner}
\end{equation}
where the sum is over all integer values of $s$ for which the
arguments of the factorials are non-negative. Later we shall use the
orthogonality relation
\begin{equation}
\int_0^\pi \sin I \rd I \,d^{\,l_1}_{n_1m_1}(I)
d^{\,l_2}_{n_2m_2}(I)=\frac{2}{2l_1+1}\delta_{m_1m_2}\delta_{n_1n_2}\delta_{l_1l_2}
\label{eq:ortho}
\end{equation}
and the symmetry relation
\begin{equation}
d^{\,l}_{-n-m}(I)=(-1)^{m-n}d^{\,l}_{nm}(I).
\label{eq:symm}
\end{equation}
With these results,
\begin{align}
\mathbfss{K}(\Lambda,\Lambda',\bfL,\bfL',\bftheta,\bftheta')=\sum_{l=0}^\infty\frac{4\pi
                    G}{2l+1}\sum_{m=-l}^l\sum_{n,n'=-l}^l
  \ri^{n-n'}y_{ln}y_{ln'}d^{\,l}_{nm}(I)d^{\,l}_{n'm}(I')\re^{\ri m(\Omega-\Omega')+in\omega-in'\omega'}\mathbfss{Q}^{\,l}_{nn'}(\Lambda,L,\Lambda',L')
\label{eq:kkred}
\end{align}
where 
\begin{equation}
\mathbfss{Q}^{\,l}_{nn'}(\Lambda,L,\Lambda',L')=-\int 
\frac{\rd\ell}{2\pi} \frac{\rd\ell'}{2\pi}\cos (nf)\cos(n'f')
\frac{r_<^l}{r_>^{l+1}}
\label{eq:qdef}
\end{equation}
and
\begin{equation}
y_{ln}\equiv Y_{ln}(\half\pi,0).
\label{eq:cln}
\end{equation}
Note that $y_{ln}=0$ unless $l-n$ is even, so the only non-zero terms
are those with $n$ and $n'$ both even if $l$ is even, or odd if $l$ is
odd.

\subsection{Mono-energetic, axisymmetric systems}

\label{sec:mono}
\noindent
We now take two steps to simplify the analysis: (i) We
restrict ourselves to mono-energetic stellar systems, in which all the
stars have the same semimajor axis $a_0$. Thus we
assume that the DF has the form
\begin{equation}
F(\Lambda,\bfL,\bftheta)=\delta(\Lambda-\Lambda_0)f(\bfL,\bftheta),\quad\mbox{with}\quad
\Lambda_0=(GM_\bullet a_0)^{1/2}. 
\label{eq:mono}
\end{equation}
To make the notation more concise, we henceforth drop $\Lambda$ as an
explicit argument of functions such as $\mathbfss{K}$ and
$\mathbfss{Q}$. (ii) We assume that the DF is axisymmetric. We have
conducted unsuccessful experiments to look for non-axisymmetric equilibria
and the simplification to axisymmetry allows a more
careful numerical exploration. Mathematically, this assumption means
that only terms with $m=0$ survive in equation (\ref{eq:kkred}) and
$f(\bfL,\bftheta)$ is independent of the angle variable $\Omega$.

After eliminating $L_z=L\cos I$ and integrating over $\Omega$, the
total mass of the stars is
\begin{equation}
M_\star=(2\pi)^2\int L \rd L\sin I \rd I \rd\omega \,f(L,I,\omega).
\label{eq:mass}
\end{equation}
The total energy may be written
\begin{align}
E&=(2\pi)^2\int L \rd L \sin I \rd I \rd\omega
 \, f(L,I,\omega)H_\mathrm{GR}(\Lambda,L) + (2\pi)^5G\int L\rd L L'\rd L'\sin I \rd I \sin I' \rd I' \rd\omega \rd\omega' f(L,I,\omega)
  f(L',I',\omega')\nonumber \\
&\quad\quad \times \sum_{l=0}^\infty\sum_{n,n'=-l}^l
  \frac{\ri^{n-n'}y_{ln}y_{ln'}}{2l+1}d^{\,l}_{n0}(I)d^{\,l}_{n'0}(I')\re^{\ri n\omega-in'\omega'}\mathbfss{Q}^{\,l}_{nn'}(L,L').
\label{eq:ered}
\end{align}
Using the relation 
\begin{equation}
Y_{ln}(I,\omega)=\left(\frac{2l+1}{4\pi}\right)^{1/2}d^{\,l}_{n0}(I)\re^{\ri n\omega}
\end{equation}
the expression for the energy simplifies to
\begin{align}
E&=(2\pi)^2\int L \rd L \sin I \rd I \rd\omega
 \, f(L,I,\omega)H_\mathrm{GR}(\Lambda,L) + 2^7\pi^6G\int L\rd L L'\rd L'\sin I \rd I \sin I' \rd I' \rd\omega \rd\omega' f(L,I,\omega)
  f(L',I',\omega')\nonumber \\
&\quad \quad \times \sum_{l=0}^\infty\sum_{n,n'=-l}^l
  \frac{\ri^{n-n'}y_{ln}y_{ln'}}{(2l+1)^2}Y_{ln}(I,\omega)Y_{ln'}^*(I',\omega')\mathbfss{Q}^{\,l}_{nn'}(L,L').
\label{eq:ered2}
\end{align}

The relative strength of relativistic precession and precession due to
self-gravity can be parametrized by
\begin{equation}
\egr\equiv
\frac{GM_\bullet}{c^2a_0}\frac{M_\bullet}{M_\star}=\frac{r_\mathrm{Sch}}{2a_0}\frac{M_\bullet}{M_\star};
\label{eq:epsgr}
\end{equation}
here $M_\star=(2\pi)^2\int L \rd L\sin I \rd I \rd\omega \,f(L,I,\omega)$ is the total mass of the stars
and $r_\mathrm{Sch}$ is the Schwarzschild radius of the black hole
(eq.\ \ref{eq:rsch}). 

We shall also use the mean eccentricity vector $\overline\bfe$ of the
system. Since the system is axisymmetric, we may assume that
$\overline\bfe$ points along the positive $z$-axis, and its value is
\begin{equation}
|\overline \bfe|\equiv \hat\bfz\,\frac{\int L \rd L\sin I \rd I \rd\omega\, f(L,I,\omega)
  e\sin I\sin\omega}{\int L \rd L\sin I \rd I \rd\omega\, f(L,I,\omega) }.
\label{eq:ebar}
\end{equation}

\subsection{Spherical systems}

\noindent
Spherically symmetric systems provide an important benchmark. In this
case the DF $f(L,I,\omega)$ depends only on $L$ and hence only terms
with $l=n=n'=0$ contribute to the energy integral. Thus
\begin{align}
E&=2^4\pi^3\int L\rd L\,f(L)H_\mathrm{GR}(\Lambda,L)+2^7\pi^6G\int L\rd L L'\rd L' f(L)f(L')\mathbfss{Q}^{\,0}_{00}(L,L').
\label{eq:esph}
\end{align}
It is straightforward to show that
\begin{equation}
\mathbfss{Q}^{\,0}_{00}(L,L')=-\int 
\frac{\rd\ell}{2\pi} \frac{\rd\ell'}{2\pi}\frac{1}{r_>}=\frac{1}{\pi^2a_0}\left[4e_>E(e_</e_>)-2e_>(1-e_<^2/e_>^2)K(e_</e_>)-\pi^2\right]
\end{equation}
where $e_<$ and $e_>$ are the smaller and larger of the eccentricities
corresponding to $L=(GM_\bullet a_0)^{1/2}(1-e^2)^{1/2}$ and $L'$, and
$E(k),K(k)\equiv \int_0^{\pi/2} \rd\theta (1-k^2\sin^2\theta)^{\pm1/2}$ are complete elliptic integrals. 

If all the stars are on circular orbits,
$f(L)\propto\delta(L-\Lambda_0)$ and
\begin{equation}
E=E_c\equiv -\frac{GM_\star^2}{a_0}\left(\half +
  3\,\egr\right).
\label{eq:ecirc}
\end{equation}
If all the stars are on radial orbits with $\egr=0$ then 
\begin{equation}
E\simeq E_r\equiv -0.29736\frac{GM_\star^2}{a_0};
\label{eq:erad}
\end{equation}
if $\egr>0$ then the energy diverges for radial orbits. 

If the DF is ergodic $f(L)$ is constant for $e<e_{\max}$ and
\begin{equation}
E=-\frac{GM_\star^2}{a_0}\times\left\{\begin{array}{ll}
0.3559 +6\,\egr & e_{\max}=1, \\
0.3703 + 4.1786\,\egr & e_{\max}=0.9. 
\end{array}\right.
\label{eq:eerg}
\end{equation}
This energy is a useful reference point because some formation
scenarios suggest, and most N-body simulations assume, that the
initial state of star clusters is close to ergodic. In the
ergodic state the DF is independent of the orbital elements other than
semimajor axis, the inverse temperature (\ref{eq:betadef}) is zero,
and the mean-square eccentricity is $\half e_{\max}^2$.

We have searched numerically for spherical maximum-entropy equilibria
outside the energy bounds set by (\ref{eq:ecirc}) and
(\ref{eq:erad}). When $\egr=0$ we have not found any, suggesting that
systems with $e=0$ and $e=e_{\max}$ have the smallest and largest
energies of any spherical systems. On the other hand, for $\egr>0$ the
systems with extreme energies may have DFs that peak at
intermediate eccentricities.

The linear stability of spherical systems is related to the existence
of lopsided maximum-entropy equilibria with the same non-Keplerian
energy. We distinguish two kinds of stability. A system is
thermodynamically stable or metastable if its entropy is a local maximum relative to
all nearby systems, spherical or non-spherical, having the same mass
and non-Keplerian energy and the same distribution of semimajor
axes. A system is dynamically stable if there are no growing modes of
the linearized collisionless Boltzmann equation. If such modes exist,
the growth time will be of order
$(a^3/GM_\bullet)^{1/2}M_\bullet/M_\star$. Since the collisionless
Boltzmann equation conserves entropy, thermodynamic stability implies
dynamical stability. If the maximum-entropy state at a given
non-Keplerian energy is lopsided, then the spherical equilibrium at
that energy must either be unstable -- a saddle point or a minimum of
the entropy -- or metastable -- a local but not global maximum of the
entropy at fixed energy.

The determination of the thermodynamic and dynamical stability of
spherical equilibria is described in Appendix \ref{sec:stable}. 

\section{Numerical methods}

\label{sec:num}

\noindent
To evaluate the entropy and energy integrals (\ref{eq:entropy}) and
(\ref{eq:ered2}) we assume that the DF $f(L,I,\omega)$ is
localized at a finite set of nodes, with eccentricities $\{e_j\}$,
$j=1,\ldots,J$, and inclinations and periapsis arguments
$\{I_k,\omega_k\}$, $k=1,\ldots,K$. Thus
\begin{equation}
  f(L,I,\omega)=\sum_{j=1}^J\sum_{k=1}^K f_{jk}\,\delta(e^2-e_j^2)\delta(\cos I-\cos
  I_k)\delta(\omega-\omega_k)
\end{equation}
where the angular momentum $L$ is related to eccentricity $e$ through
$L^2=\Lambda_0^2(1-e^2)$. The mass associated with node $(j,k)$ is given
by  (cf.\ eq.\ \ref{eq:mass})
\begin{equation}
  M_{jk}=2\pi^2 \Lambda_0^2 f_{jk}.
\end{equation}
The entropy (\ref{eq:entropy}) is approximated as
\begin{equation}
  S=-\sum_{j=1}^J\sum_{k=1}^K M_{jk}\log (M_{jk}/V_{jk})
  \label{eq:entdef}
\end{equation}
where $V_{jk}$ is the phase-space volume associated with node $(j,k)$,
which we determine below. 

Equation (\ref{eq:ered2}) for the energy becomes
\begin{align}
E&=-6\pi^2\frac{(GM_\bullet)^3}{c^2a_0}\sum_{j=1}^J\sum_{k=1}^K\frac{f_{jk}}{(1-e_j^2)^{1/2}}
  + 2^5\pi^6 G^3M_\bullet^2a_0^2 \\
&\quad\times \sum_{j,j'=1}^J\sum_{k,k'=1}^K f_{jk}f_{j'k'}\sum_{l=0}^\infty\sum_{n,n'=-l}^l
  \frac{\ri^{n-n'}y_{ln}y_{ln'}}{(2l+1)^2}Y_{ln}(I_k,\omega_k)Y_{ln'}^*(I_k',\omega_k')\mathbfss{Q}^{\,l}_{nn'}(e_j,e_{j'});\nonumber
\end{align}
we have changed the arguments of $\mathbfss{Q}^{\,l}_{nn'}$ from
$L=\Lambda_0(1-e^2)^{1/2}$ to eccentricity $e$, with a similar change
from $L'$ to $e'$.

We truncate the sum over $l$ at some maximum value $l_\mathrm{max}$. The sums
can be shortened to sums over non-negative $n$ by observing that
$\mathbfss{Q}^{\,l}_{nn'}$ is even in both $n$ and $n'$, that
$y_{l-n}=(-1)^ny_{ln}$, and that
$Y_{l-n}(I,\omega)=(-1)^nY_{ln}^*(I,\omega)$. Thus
\begin{align}
\label{eq:ered4}
E&=-6\pi^2\frac{(GM_\bullet)^3}{c^2a}\sum_{j=1}^J\sum_{k=1}^K\frac{f_{jk}}{(1-e_j^2)^{1/2}}
 +2^7\pi^6G^3M_\bullet^2a_0\sum_{j,j'=1}^J\sum_{k,k'=1}^K f_{jk}f_{j'k'}\\
&\times \sum_{l=0}^{l_\mathrm{max}}\sum_{n,n'=0}^l
  c_{ln}c_{ln'} Y_{ln}(I_k,0)Y_{ln'}(I_{k'},0)\cos n(\omega_k+\half\pi) \cos n'(\omega_{k'}+\half\pi)\,\mathbfss{Q}^{\,l}_{nn'}(e_j,e_{j'}).\nonumber
\end{align}
where 
\begin{equation}
c_{ln}\equiv \frac{y_{ln}}{(1+\delta_{n0})(2l+1)}.
\label{eq:cdef}
\end{equation}
Since the second term in (\ref{eq:ered4}) is symmetric in the primed and
unprimed variables the summation over $(j,k)$ and $(j',k')$ can be shortened by roughly a
factor of two, yielding further savings in the computation time. 

We use a set of nodes $\{e_j\}$ that are uniformly distributed in $e^2$,
\begin{equation}
  e_j^2=\frac{(j-\half)}{J}e_{\max}^2, \quad j=1,\ldots,J, 
\end{equation}
where $e_{\max}$ is defined in \S\ref{sec:gr}.  The nodes in
inclination and periapsis argument are assigned to a Lebedev
quadrature grid. A Lebedev grid $\{\theta_k,\phi_k,w_k\}$ of order $P$
is a Gaussian quadrature algorithm in the sense that
\begin{align}
&\int_0^\pi \sin\theta \rd\theta\int_0^{2\pi}\rd\phi\,
h(\sin\theta\cos\phi,\sin\theta\sin\phi,\cos\theta) \nonumber \\
&\quad = 4\pi \sum_{k=1}^K w_k
  h(\sin\theta_k\cos\phi_k,\sin\theta_k\sin\phi_k,\cos\theta_k) 
\label{eq:lebdef}
\end{align}
is exact when the function $h(x,y,z)$ is any polynomial of order
$\le P$. The weights $w_k$ sum to unity and the number of points $K$
is related to the order $P$; for example, for $P=5,7,9,11,13,15$ we
have $K=14,26,38,50,74,86$. We have checked that assigning the angular
nodes to a Cartesian grid in $\cos I\in[0,\pi]$ and
$\omega\in[0,2\pi)$ yields the same results, although with lower
accuracy for the same number of grid points.

The phase-space volume enclosed by a surface $S$ is
\begin{equation}
 V_S= \int_S \rd\Lambda \rd L \rd L_z \rd\omega \rd\Omega \rd\ell=\half \int_S \Lambda^2
  \rd\Lambda \rd e^2 \sin I \rd I \rd\omega \rd\Omega \rd\ell.
\end{equation}
The integrals over $\Lambda$, $\ell$, and $\Omega$ are the same for all stars (because 
we consider mono-energetic, orbit-averaged, axisymmetric systems), so
if we evaluate the remaining three integrals using the method we have
described above we find
\begin{equation}
  V_S=\mbox{constant}\times \sum_{j=1}^J \sum_{k=1}^K W(j,k,S) w_k
\end{equation}
where $W(j,k,S)$ is 1 if $(e_j,I_k,\omega_k)$ is inside $S$ and zero
otherwise. Thus the volume associated with node $(j,k)$ is
\begin{equation}
  V_{jk}=\mbox{constant}\times w_k.
\end{equation}
Since $V_{jk}$ is only used in the expression for the entropy
(\ref{eq:entdef}) and here it appears only in the argument of a
logarithm, we can set the constant to unity if we assume that the
entropy is only defined to within a constant. 

The matrix $\mathbfss{Q}^{\,l}_{nn'}(e_j,e_j')$ (eq.\ \ref{eq:qdef})
is computed once and for all at the start. The most convenient
integration variable for this task is the eccentric anomaly $u$, which
is related to the variables in equation (\ref{eq:qdef}) by
$\ell=u-e\sin u$, $r=a(1-e\cos u)$, and
$\cos f=(\cos u-e)/(1-e\cos u)$.

We then maximize the entropy (\ref{eq:entdef}) subject to the
non-linear constraint that the energy (\ref{eq:ered4}) is fixed and the
linear constraints that the total mass $\sum_{jk}M_{jk}=1$ and
$M_{jk}\ge0$\footnote{The total angular momentum, which we assume to
  be zero, is also conserved. This constraint is automatically
  satisfied by our solutions, because the energy (\ref{eq:ered4}) is
  invariant when orbit directions are reversed. Thus maximum-entropy
  solutions have equal numbers of orbits going in opposite
  directions.}. The initial conditions for the optimization algorithm
are chosen in one of two ways: (i) the periapsis directions are
restricted to lie within $45\deg$ of the positive $z$-axis
($\sin I_j\sin\omega_j > 2^{-1/2}$); this encourages the routine to
find lopsided states if they exist; (ii) if we are finding a sequence
of equilibria, say for a set of energies $E_n$, the equilibrium for
energy $E_n$ is used as the initial state when seeking the equilibrium
for energy $E_{n+1}$.

We use the optimization routine E04UCF from the NAG (Numerical
Algorithms Group) library. Most of the computation time is spent on
evaluating the sum (\ref{eq:ered4}) but this task is easy to
parallelize. 

The accuracy of the calculations depends on the number of eccentricity
grid points ($J$), the number of angular grid points ($K$), and the
maximum multipole $l_\mathrm{max}$. Increasing $J$ and
$K$ improves the accuracy of the calculations but the required
computing time grows rapidly with $J$ and $K$; moreover as the number
of variables $JK$ grows it becomes more difficult for the optimization
routine to converge.  The values
we use are a compromise between these conflicting demands: typically
$l_\mathrm{max}=8$, $J=16$, and $K=50$ (corresponding to order $P=11$). In this
case we have an optimization problem with
800 variables.

\section{Results}

\label{sec:results}

\begin{figure}
\includegraphics[width=\textwidth]{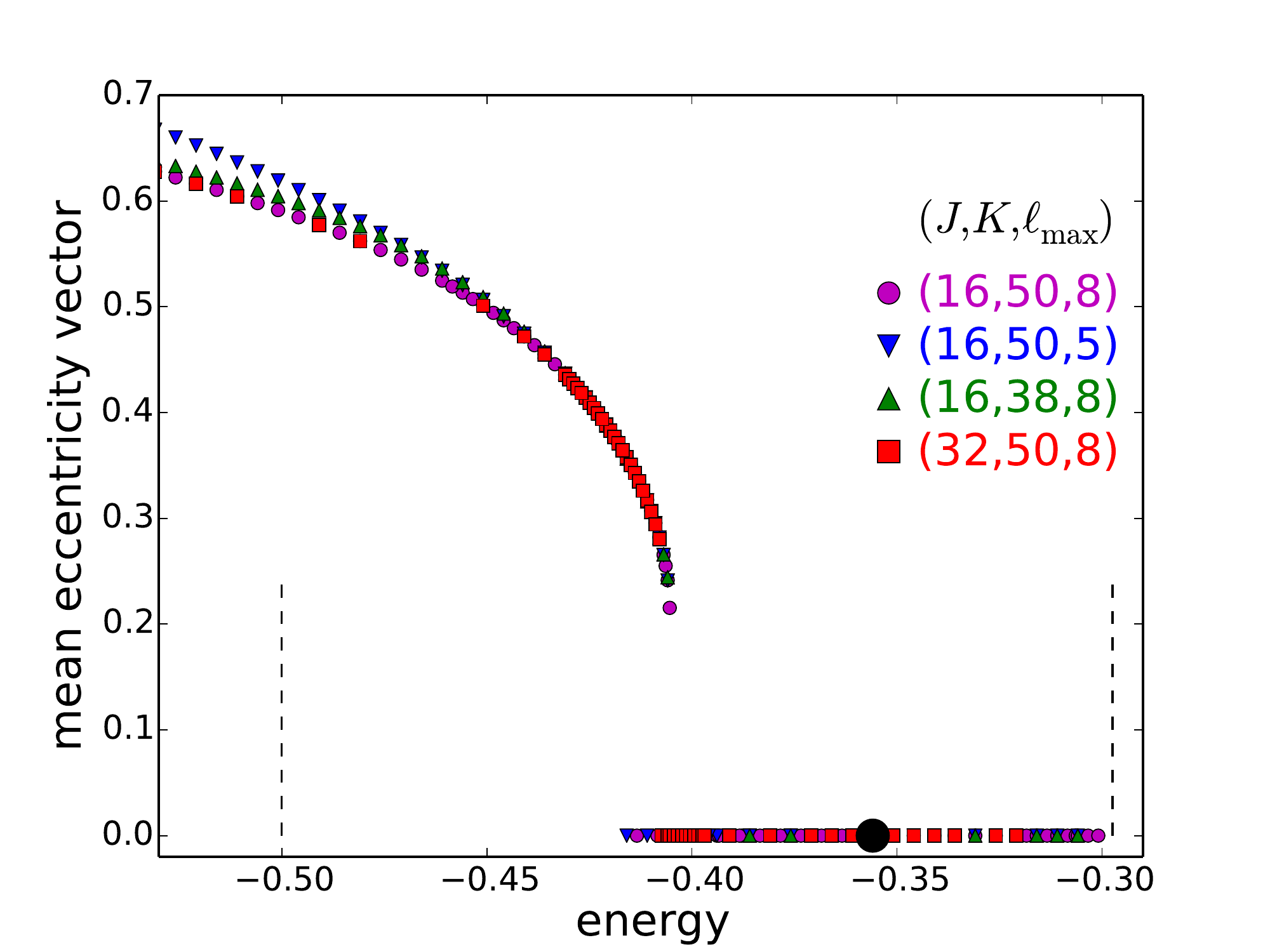}
\caption{\small Magnitude of the mean eccentricity vector in a
  maximum-entropy mono-energetic stellar system, as a function of the
  self-gravitational energy measured in units of $GM_\star^2/a_0$
  (eq.\ \ref{eq:ered4}). Relativistic precession is assumed to be
  negligible and the maximum allowed eccentricity is unity ($\egr=0$,
  $e_{\max}=1$). The ergodic DF (inverse temperature $\beta=0$) is
  marked by a black circle at $E=-0.3559$. Note that there is a small
  interval ($-0.42 \lesssim E \lesssim -0.41$) in which both spherical and lopsided
  systems are local entropy maxima. Each set of colored symbols
  represents a specific combination of the number of eccentricity grid
  points ($J$), the number of points in the Lebedev quadrature ($K$), and the
  maximum multipole $l_\mathrm{max}$, shown in the legend as $(J,K,l_\mathrm{max})$. The
  dashed vertical lines denote the minimum and maximum energy
  of spherical equilibria, $E_c =-0.5$ and $E_r=-0.2974$ (cf.\ eqs.\
  \ref{eq:ecirc}--\ref{eq:eerg}).}
\label{fig:one}
\end{figure}

\begin{figure}
\includegraphics[width=\textwidth]{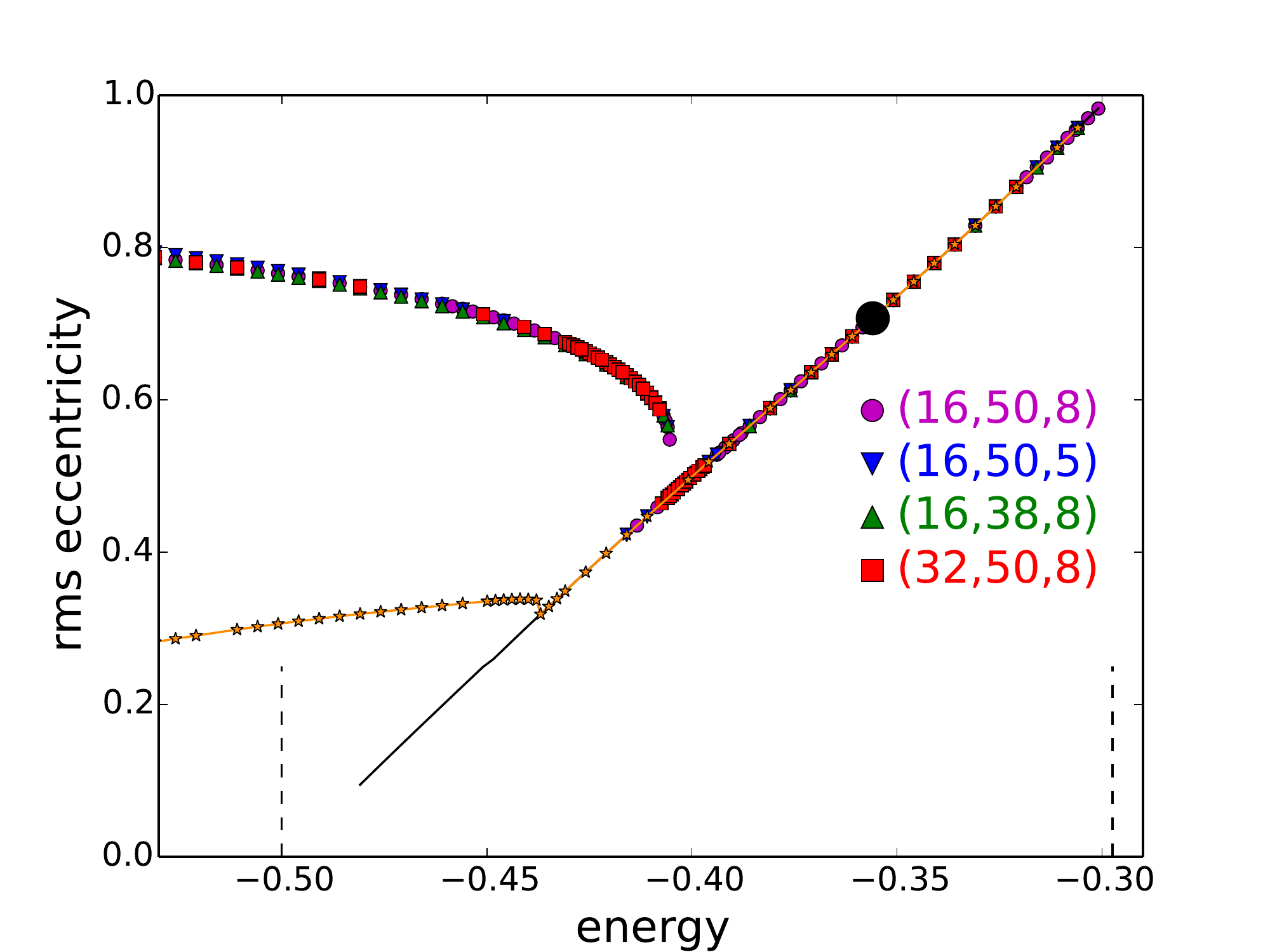}
\caption{\small Root mean square (rms) eccentricity in a
  maximum-entropy mono-energetic stellar system, as a function of the 
  self-gravitational energy $E$ measured in units of $GM_\star^2/a_0$. The
  symbols have the same meaning as in Figure
  \ref{fig:one}. The solid line shows the rms eccentricity for
  spherical models. The ergodic system (black circle) has rms
  eccentricity $2^{-1/2}=0.7071$. The orange line with star symbols shows
  maximum-entropy models in which the mean eccentricity vector is
  constrained to be zero. }
\label{fig:two}
\end{figure}

\subsection{No relativistic precession}

\label{sec:norel}

\noindent
We first examine the properties of maximum-entropy states in the
absence of relativistic precession ($\egr=0$).  Figure \ref{fig:one}
shows the mean eccentricity vector (eq.\ \ref{eq:ebar}) as a function
of the self-gravitational energy of the stars measured in units of
$GM_\star^2/a_0$. In these units spherically symmetric systems have
energies between the vertical dashed lines at $E=-0.5$ (circular
orbits, eq.\ \ref{eq:ecirc}) and $-0.2974$ (radial orbits, eq.\
\ref{eq:erad}). Ergodic systems have $E=-0.3559$ (eq.\ \ref{eq:eerg})
and are marked by a filled black circle. The colored symbols represent
different choices for the integration parameters $J$, $K$, and
$l_\mathrm{max}$. All of the choices shown yield similar results even
though the number of eccentricity grid points $J$ ranges from 16 to
32, the number of angular grid points $K$ ranges from 38 to 50 ($P$
between 9 and 11), and the maximum order of the spherical-harmonic
expansion $l_\mathrm{max}$ ranges from 5 to 8. Thus we are confident
that the numerical methods have converged reasonably well.

Equilibrium systems with energy $E > -0.41$ have zero mean
eccentricity vector and are spherically symmetric\footnote{For
  spherical symmetry it is necessary but not sufficient that the
  eccentricity vector vanishes. We have checked explicitly that for
  $E > -0.41$ the DF $f_{jk}$ is independent of the angular node $k$
  at given eccentricity $e_j$.}. For $E <-0.41$ the maximum-entropy
systems acquire a non-zero mean eccentricity vector, indicating that
they are lopsided. In the language of condensed-matter physics, there
is an order-disorder phase transition as the stellar system is cooled.

\begin{figure}
\includegraphics[width=\textwidth]{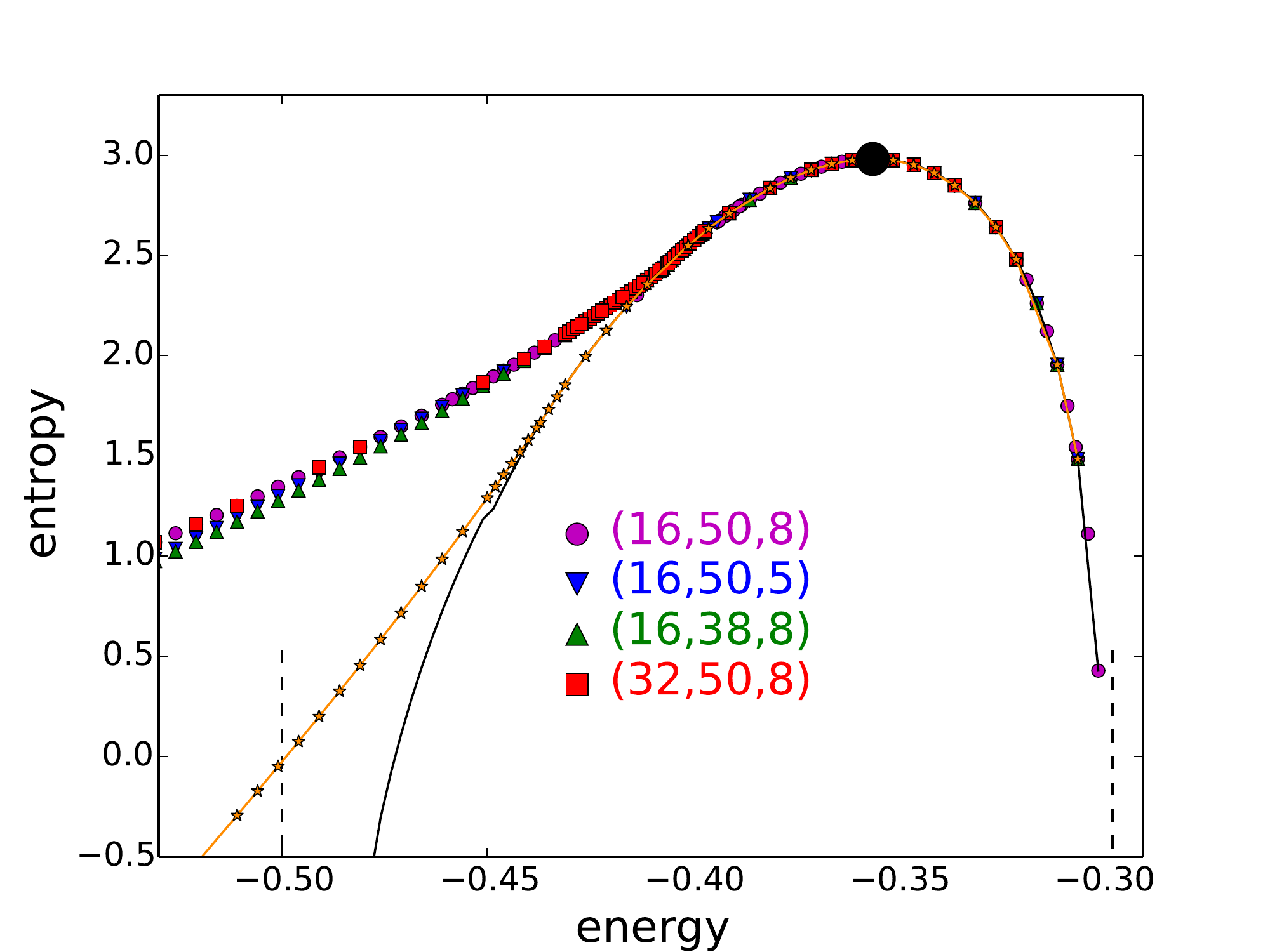}
\caption{\small Entropy of a mono-energetic stellar system as a
  function of the self-gravitational energy $E$. The entropy has an
  arbitrary additive constant. Symbols have the same meaning as in
  Figure \ref{fig:two}.}
\label{fig:three}
\end{figure}

Figures \ref{fig:two} and \ref{fig:three} show the rms eccentricity
and entropy of these models as a function of energy. The
solid black curve in each figure is the rms eccentricity or entropy of
the maximum-entropy spherical model.  In Figure \ref{fig:two}, the rms
eccentricity of the spherical model grows smoothly
from 0 to 1 as the energy grows from its minimum value for spherical
systems, $-0.5 GM_\star^2/a_0$, to its maximum of
$-0.2974\,GM_\star^2/a_0$. 

In Figure \ref{fig:three}, the slope $dS/dE = \beta$ is negative for
energies larger than that of the ergodic state (dotted line at
$E=-0.3559$), which means that the equilibrium temperature $1/\beta$
is negative. 

In both Figure \ref{fig:two} and \ref{fig:three}, the black curves
coincide with the colored markers for $E> -0.41$, confirming
that in this region the maximum-entropy equilibrium is spherical. For
$E < -0.41$ the maximum-entropy spherical states have smaller
entropy than the lopsided states, which are global entropy maxima.

Some of this behaviour can be illuminated by examining the linear
stability of spherical maximum-entropy systems using the methods of
Appendix \ref{sec:stable}. When $l=1$ and relativistic precession is absent,
it can be shown analytically that spherical equilibria are always
dynamically stable or at least neutrally stable
\citep{tre05,pps07}. Numerical solutions of the eigenvalue equation for the
matrix $\mathbfss{V}^{\,l}$ (eq.\ \ref{eq:vdef}) over the range of
energies $-0.310$ to $-0.489$ show that the spherical equilibria are
also dynamically stable for $l=2,3,\ldots,8$.  Numerical solutions of
the eigenvalue equation for $\mathbfss{R}^{\,l}$ (eq.\
\ref{eq:discrete}) show that they are also thermodynamically stable
for odd values of $l=1,3,5,7$. However, they can be thermodynamically
unstable for even values of $l$: for $l=2$ the spherical system is
unstable when $E<-0.439$, for $l=4$ when $E<-0.478$, for $l=6$ when
$E<-0.488$, and for $l=8$ when $E<-0.491$.

Since spherical systems are dynamically and thermodynamically stable
(at least up to $l=8$) for energy $E>-0.439$, the phase transition at
larger energies must arise because the spherical equilibrium is metastable,
i.e., it is a local entropy maximum but not a global one. This
conclusion is consistent with the observation that our numerical
optimization algorithm finds both spherical and lopsided
maximum-entropy states for energies in the range $E=-0.405$ to
$E=-0.418$. In principle there are spherical maximum-entropy states up
to the onset of the $l=2$ instability at $E=-0.439$ but the maxima are
very shallow, and difficult to detect without higher resolution simulations.

The orange curve in Figures \ref{fig:two} and \ref{fig:three} shows
the maximum-entropy state when the mean eccentricity vector is
constrained to be zero. For $E > -0.439$ the maximum-entropy state
with $\overline \bfe=0$ is spherical, so the orange and black lines
coincide. For $E<-0.439$ the spherical state is thermodynamically
unstable to $l=2$ perturbations, which leave the eccentricity vector
unchanged, so the maximum-entropy state with $\overline\bfe=0$ is
non-spherical (zero dipole but non-zero quadrupole moment), and the
orange line has higher entropy than the black line.

\subsection{Effects of relativistic precession}

\begin{figure}
\includegraphics[width=\textwidth]{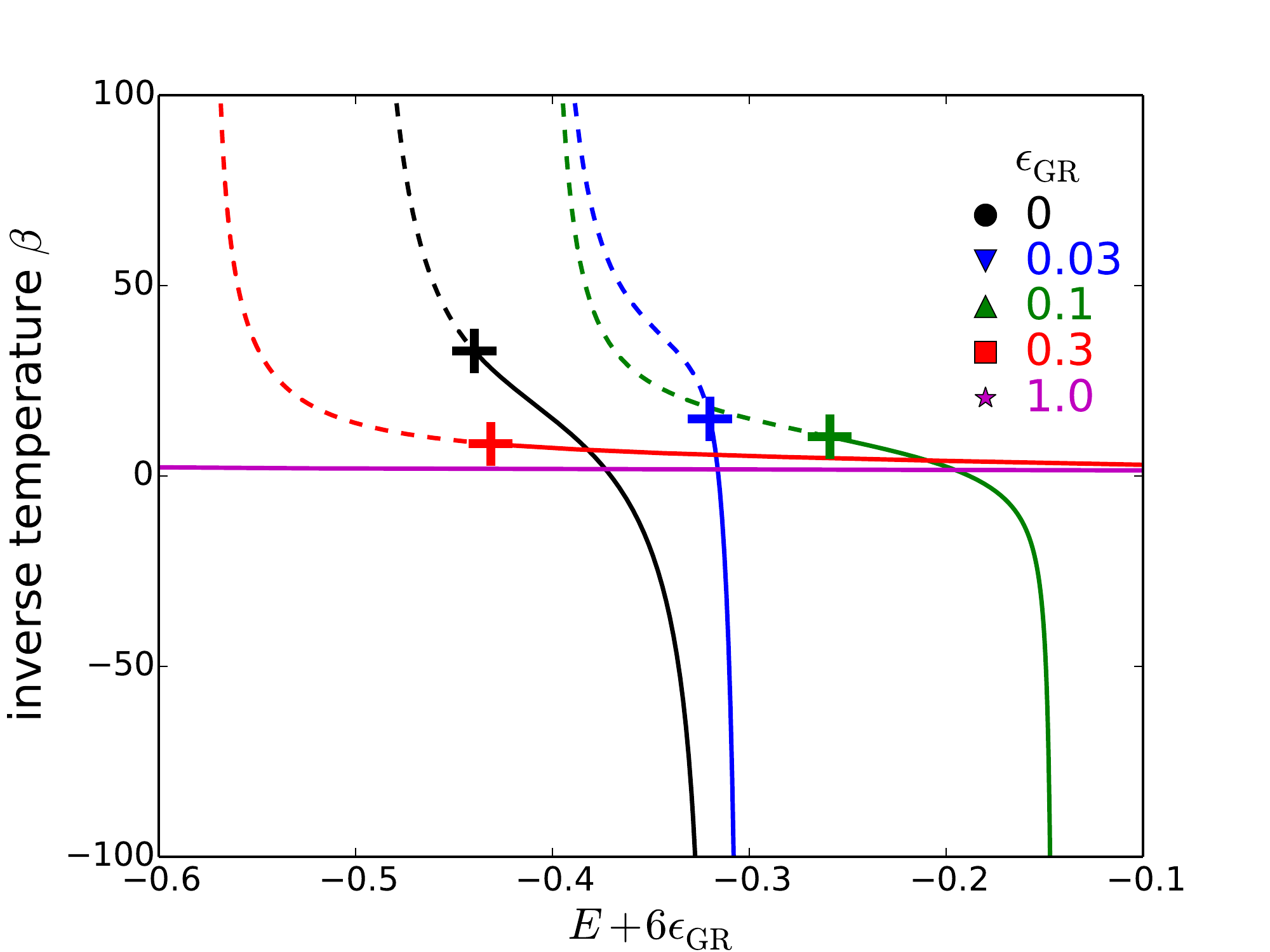}
\caption{\small Inverse temperature in a maximum-entropy
  mono-energetic stellar system, as a function of the non-Keplerian
  energy $E$ measured in units of $GM_\star^2/a_0$.  Each curve
  represents a different value of the parameter $\egr$, which measures
  the strength of relativistic precession relative to precession from
  self-gravity (eq.\ \ref{eq:epsgr}). On the horizontal axis, the
  energy $E$ is offset by $6\egr$ to help all of the curves to fit on
  a single figure. The solid curves denote stable equilibria while the
  dashed curves denote unstable equilibria; the boundary between
  stability and instability is marked on each curve by a plus
  sign. The most unstable mode is an $l=2$ thermodynamic instability
  for $\egr=0$ and an $l=1$ dynamical instability for $\egr>0$. }
\label{fig:seven}
\end{figure}

\begin{figure}
\includegraphics[width=\textwidth]{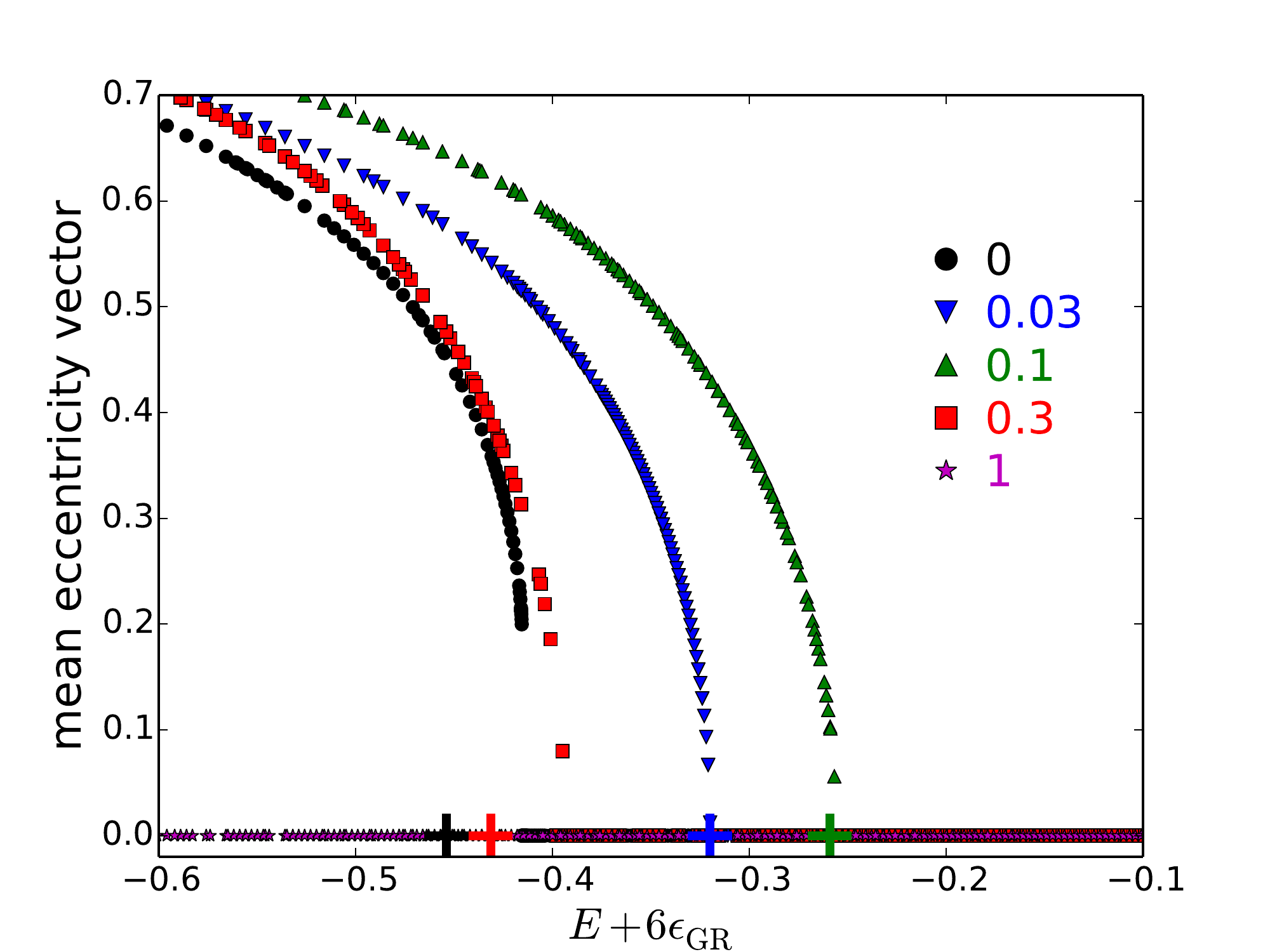}
\caption{\small Magnitude of the mean eccentricity vector in a
  maximum-entropy mono-energetic stellar system, as a function of the
  non-Keplerian energy $E$ measured in units of $GM_\star^2/a_0$.
  Each set of colored symbols represents a different value of the
  parameter $\egr$, which measures the strength of relativistic
  precession relative to precession from self-gravity (eq.\
  \ref{eq:epsgr}). On the horizontal axis,
  the energy $E$ is offset by $6\egr$ to help all of the curves to fit
  on a single figure. All spherical systems to the left of
  the plus signs for $\egr=0.03,0.1,$ and $0.3$ are subject to $l=1$
  thermodynamic and dynamical instabilities. The models
  shown by symbols have numerical grid $(J,L,l_\mathrm{max})=(16,11,8)$.}
\label{fig:four}
\end{figure}

\begin{figure}
\includegraphics[width=\textwidth]{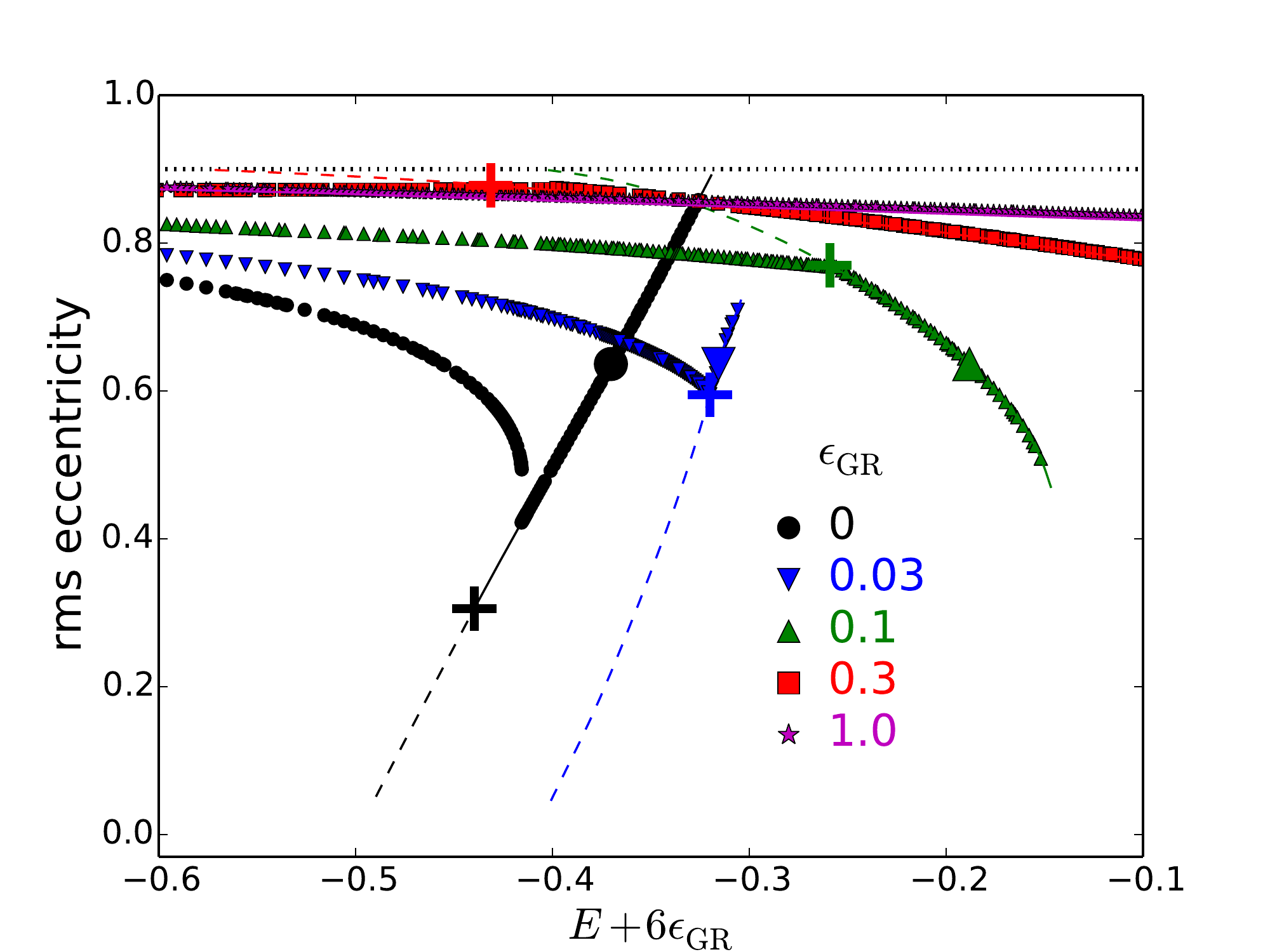}
\caption{\small Root mean square (rms) eccentricity in a
  maximum-entropy mono-energetic stellar system, as a function of the
  non-Keplerian energy. The symbols have the same meaning as in Figure
  \ref{fig:four}. The colored lines show the rms eccentricity for
  maximum-entropy spherical models; these are stable to the right of
  the plus symbols (solid lines) and unstable to the left (dashed
  lines). The expanded symbols denote ergodic systems, which all have rms
  eccentricity 0.6708. The phase space for these systems has maximum
  eccentricity $e_{\max}=0.9$, denoted by the horizontal dotted
  line. }
\label{fig:five}
\end{figure}

\begin{figure}
\includegraphics[width=\textwidth]{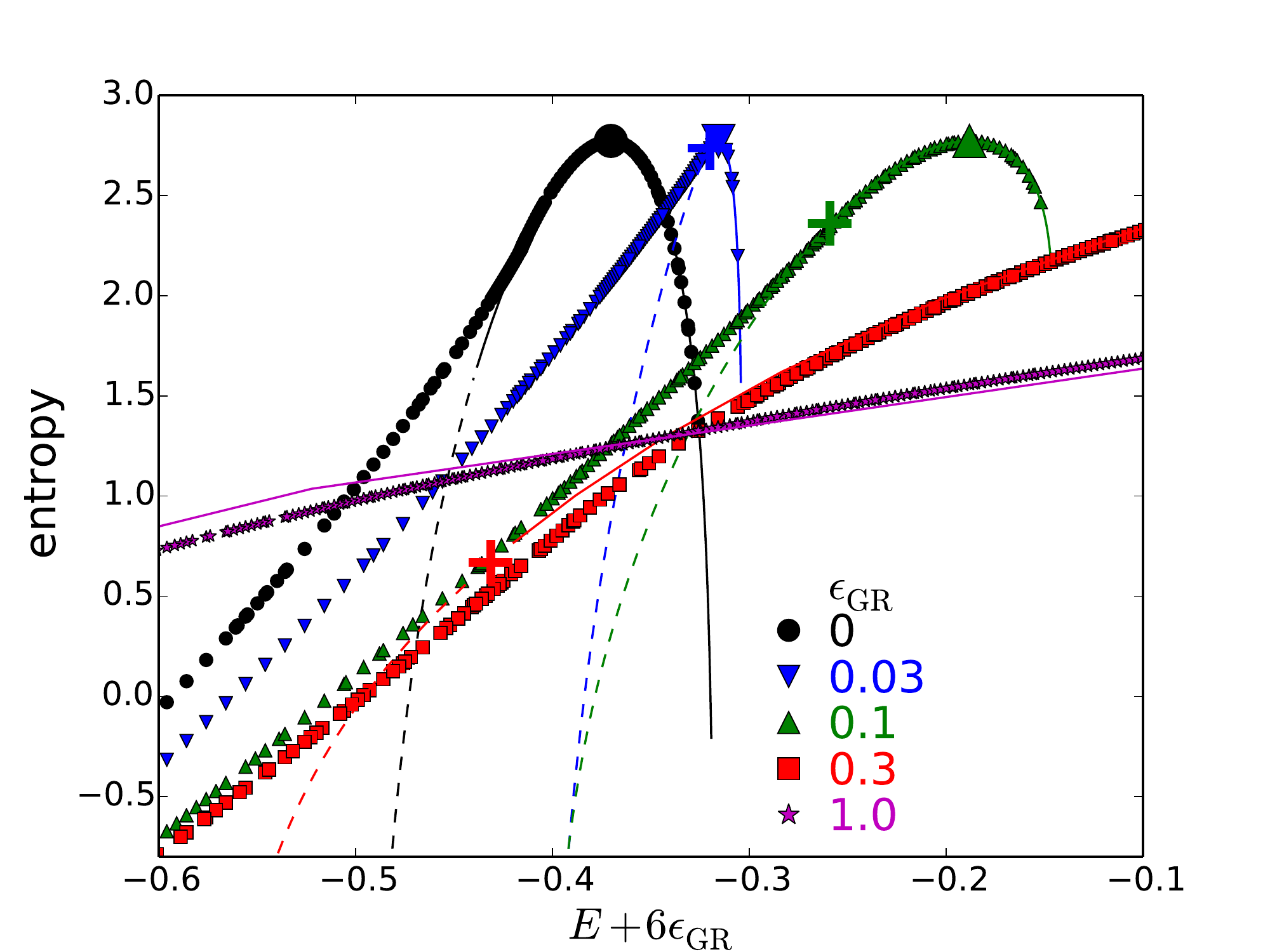}
\caption{\small Maximum entropy of a mono-energetic stellar system, as
  a function of the non-Keplerian energy $E$, measured in units
  of $GM_\star^2/a_0$. The symbols have the same meaning as in Figures
  \ref{fig:four} and \ref{fig:five}. The small offset between the
  symbols and the solid curve for $\egr=0.3$ and 1 probably reflect
  the limited eccentricity resolution of the models ($J=16$), which
  means that they are inaccurate when the rms eccentricity is close to
  unity.}
\label{fig:six}
\end{figure}

\noindent
We parametrize the importance of relativistic precession compared to
self-gravity by $\egr$, defined in equation (\ref{eq:epsgr}). To avoid
divergences in the Hamiltonian when $\egr$ is non-zero, we cut
off the DF above some maximum eccentricity $e_{\max}$, which we
arbitrarily set to be $e_{\max}=0.9$. In our plots we use $E+6\egr$ as
the independent variable. Here $E$ is the non-Keplerian energy of the
stellar system (eq.\ \ref{eq:ered4}) including both the
self-gravitational energy and the energy from the relativistic
Hamiltonian (\ref{eq:hgr}), measured in units of $GM_\star^2/a_0$. The
term $6\egr$ is an empirical offset, introduced solely to enable us to
plot systems with quite different energies on the same figure.

Figure \ref{fig:seven} shows the inverse temperature of spherical
systems as a function of energy, for several values of the
relativistic parameter $\egr$. The inverse temperature declines with
increasing energy, so all of these systems have positive heat
capacity. The plus signs separate stable systems (solid lines) from
unstable ones (dashed lines). For $\egr=0$ the transition occurs at
$E=-0.454$ through an $l=2$ thermodynamic instability (to be compared
to $E=-0.439$ for the system examined in \S\ref{sec:norel}, which had
$e_{\max}=1$ compared to $e_{\max}=0.9$). For $\egr>0$ the transition
occurs through an $l=1$ dynamical instability.

Figure \ref{fig:four}, the analog to Figure \ref{fig:one}, plots the
mean eccentricity vector as a function of energy. The order-disorder
phase transition is present for all five values of the relativistic
parameter $\egr$, although the transition is off the figure (at
$E=-7.071$) for $\egr=1$. For energies below the phase transition the
maximum-entropy equilibria are lopsided, while above the phase
transition they are spherical.

The nature of the phase transition depends on the strength of the
relativistic effects. As in Figure \ref{fig:one}, when $\egr=0$ (i)
spherically symmetric states are metastable entropy maxima for
energies just below (or inverse temperatures above) the phase
transition at $E\simeq -0.41$; (ii) eventually, as the energy
declines, an $l=2$ thermodynamic instability sets in; (iii) spherical
systems are always dynamically stable and always stable to $l=1$
disturbances. In contrast, when $\egr>0$ spherical systems with
sufficiently low energy are thermodynamically and dynamically unstable
to $l=1$ disturbances. As $\egr$ grows the onset of this instability,
marked by colored plus signs in Figure \ref{fig:four}, shifts closer
and closer to the phase transition. Eventually the onset
of the $l=1$ instability coincides with the phase transition so there
is no metastable spherical state at energies below the transition
energy.

Figure \ref{fig:five} and \ref{fig:six} show the rms eccentricity and
entropy for the same systems. When $\egr$ is small, the rms
eccentricity increases as the energy grows (or inverse temperature
declines). However, between $\egr=0.03$ and $\egr=0.1$ this behaviour
reverses, and the rms eccentricity declines as the energy grows for
spherical systems. Below this transition the rms eccentricity of the
lopsided equilibrium is larger than that of the spherical equilibrium
with the same energy; above the transition this ordering is reversed.

\section{Discussion}

\label{sec:disc}

\noindent
We have shown that a phase transition from spherical to lopsided
equilibria occurs in an idealized model of a black-hole star cluster
dominated by a central black hole. We now ask what conditions are
needed for this transition to be present in realistic clusters.

\subsection{A simplified dynamical model of a black-hole star cluster}

\begin{figure}
\includegraphics[width=\textwidth]{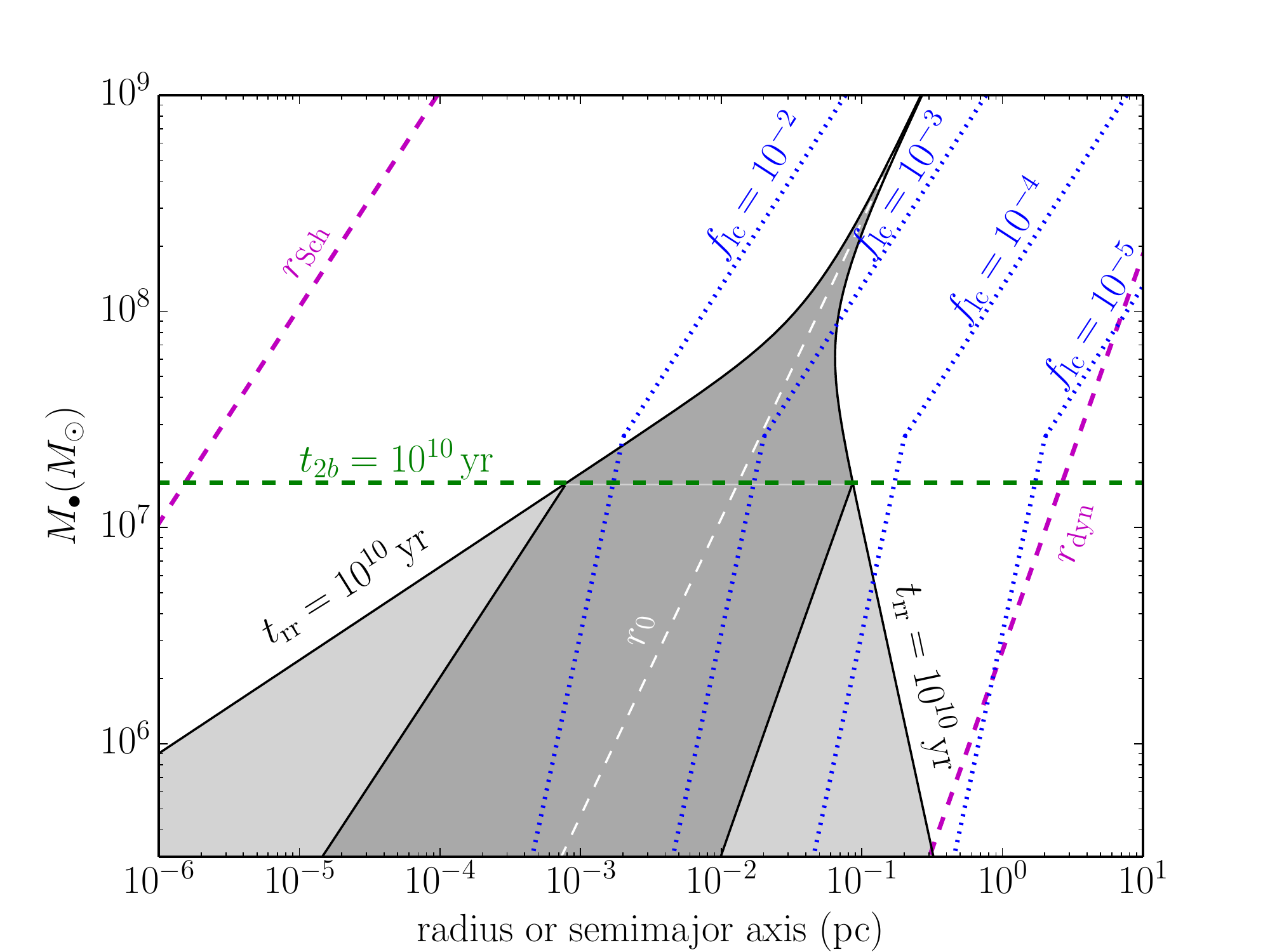}
\caption{\small Properties of the model of a black-hole star cluster
  described in \S\ref{sec:simple}. The horizontal axis is radius or
  semimajor axis and the vertical axis is the mass of the central
  black hole. The Schwarzschild radius of the black hole
  $r_\mathrm{Sch}$ (eq.\ \ref{eq:rsch}) and the dynamical radius
  $r_\mathrm{dyn}$ at which the enclosed stellar mass equals the
  black-hole mass (eq.\ \ref{eq:rdyn}) are shown as dashed magenta
  lines. Above the horizontal dashed green line the two-body
  relaxation time $t_{2b}$ (eq.\ \ref{eq:t2b}) exceeds the typical
  galaxy age of $10^{10}\yr$. Within the shaded region the resonant
  relaxation time $t_\mathrm{rr}$ (eq.\ \ref{eq:t2b}) is less than
  $10^{10}\yr$ (at a typical eccentricity $e=0.5$), and in the dark
  shaded region $t_\mathrm{rr}$ is also less than the two-body time
  $t_{2b}$. The dashed white line is the locus of $r_0$ for $e=0.5$
  (eq.\ \ref{eq:r0}). Apsidal precession is dominated by relativity to
  the left of this line and by the stellar self-gravity to the right,
  and on this line the relativistic parameter is
  $\egr=0.1666$. Finally, the dotted blue lines mark loci of constant
  $f_\mathrm{lc}$, the fraction of phase space occupied by the loss
  cone (eq.\ \ref{eq:flc}). All curves assume an effective stellar
  mass $m_\mathrm{eff}=\langle m^2\rangle/\langle m\rangle= 1\msun$.
}
\label{fig:eight}
\end{figure}

\label{sec:simple}

\noindent
We consider a cluster containing a central black hole of mass
$M_\bullet$. The Schwarzschild radius of the black hole is 
\begin{equation}
r_\mathrm{Sch} \equiv \frac{2GM_\bullet}{c^2}=9.57\times
10^{-6}\pc\,\frac{M_\bullet}{10^8\msun}.
\label{eq:rsch}
\end{equation}
The black hole is surrounded by a spherical star
cluster. The mass of stars interior to radius $r$ is $M_\star(r)$ and
we define the dynamical radius of the black hole, $r_\mathrm{dyn}$, by
$M_\star(r_\mathrm{dyn})=M_\bullet$ \citep{bt08}.  For
$r\la r_\mathrm{dyn}$ the gravitational force is dominated by the
central black hole, and the orbital angular frequency of a star with
semimajor axis $a$ is 
\begin{equation}
\Omega^{-1}(a)=\left(\frac{a^3}{GM_\bullet}\right)^{1/2}=1.49\times10^3\yr\left(\frac{a}{\pc}\right)^{3/2}
\left(\frac{10^8\msun}{M_\bullet}\right)^{1/2}.
\label{eq:omdef}
\end{equation}

The mass of stars inside radius $r$ is assumed to be a power law,
\begin{equation}
M_\star(r)=M_\bullet\left(\frac{r}{r_\mathrm{dyn}}\right)^{3-\gamma}.
\end{equation} 
For numerical calculations we set $\gamma=1.5$, a typical value for black-hole star
clusters. Assuming that the velocity-dispersion tensor of the cluster is
isotropic and solving the Jeans equation for the one-dimensional
velocity dispersion $\sigma(r)$ (eq.\ 4.216 in \citealt{bt08}) we find
that  
\begin{equation}
\sigma(r)=\left[\frac{GM_\bullet}{(1+\gamma)r}\right]^{1/2}=415{\,\mbox{km\,s}^{-1}}\left(\frac{M_\bullet}{10^8\msun}\frac{1\pc}{r}\right)^{1/2},\quad
r\ll r_\mathrm{dyn}. 
\end{equation}
We assume that the dispersion at the dynamical radius,
$\sigma(r_\mathrm{dyn})$, is equal to the dispersion of the central
region of the galaxy outside the dynamical radius, which we denote
$\sigma_\mathrm{g}$. In early-type galaxies $\sigma_\mathrm{g}$ is strongly correlated
with the central black-hole mass and this correlation can be
approximated as \citep[e.g.,][]{kh13}
\begin{equation}
M_\bullet \simeq 3.1\times 10^8\msun\left(\frac{\sigma_\mathrm{g}}{200\,\mbox{km\,s}^{-1}}\right)^{4.4}.
\end{equation}

With these assumptions the local properties of the cluster are fully
described by two parameters, the black-hole mass and the radius or
semimajor axis. The dynamical radius can be written
\begin{equation}
r_\mathrm{dyn}=\frac{GM_\bullet}{2.5\sigma_\mathrm{g}^2} =
7.2\pc\left(\frac{M_\bullet}{10^8\msun}\right)^{0.545}.
\label{eq:rdyn}
\end{equation}
The mass density of stars at radii $r\ll r_\mathrm{dyn}$ is 
\begin{equation}
\rho(r)=\frac{1}{4\pi r^2}\frac{d
  M_\star}{dr}=\frac{3M_\bullet}{8\pi (r_\mathrm{dyn}r)^{3/2}}
=6.19\times10^5\msun\pc^{-3} 
\left(\frac{M_\bullet}{10^8\msun}\right)^{0.182}\left(\frac{1\pc}{r}\right)^{1.5}.
\end{equation}

We generalize the relativistic parameter $\egr$  (eq.\ \ref{eq:epsgr})
to a function of radius, 
\begin{equation}
\egr(r)=\frac{GM_\bullet^2}{c^2rM_\star(r)}=\frac{r_\mathrm{Sch}r^{1.5}_\mathrm{dyn}}{2r^{2.5}}=
9.23\times10^{-5}\left(\frac{M_\bullet}{10^8\msun}\right)^{1.818}\left(\frac{1\pc}{r}\right)^{2.5}.
\end{equation}

The apsidal precession rate for an orbit of semimajor axis $a$ and
eccentricity $e$ is
\begin{align}
\dot\omega(a,e)&=\Omega(a)\left(\frac{a}{r_\mathrm{dyn}}\right)^{1.5}w(e)
+\frac{3\Omega(a) GM_\bullet}{c^2a(1-e^2)} \nonumber \\
&=\Omega(a)\left(\frac{a}{r_\mathrm{dyn}}\right)^{1.5}\left[w(e)+ \frac{3\egr(a)}{1-e^2}\right],
  \quad r\ll r_\mathrm{dyn}.
\end{align}
Here 
\begin{equation}
w(e)=\frac{2(1+e)(1-e)^{1/2}}{\pi e^2}\left[(1-e)K(k)-E(k)\right],
\quad k^2=\frac{2e}{1+e},
\end{equation}
where $K$ and $E$ are complete elliptic integrals. The function $w(e)$
is negative (i.e., the precession due to self-gravity is retrograde) for all eccentricities $e$ between 0 and 1;
$w(0)=-\frac{3}{4}$, $w(0.5)=-0.66634$, and $w(e)\to
-4(1-e)^{1/2}\pi$ as $e\to 1$. 

The two-body relaxation time is given by equation
(7.106) of \citet{bt08}, 
\begin{align}
t_\mathrm{2b}&=0.34\,\frac{\sigma^3}{G^2m_\mathrm{eff}\rho\log\Lambda} 
\\
&=1.38\times
  10^{11}\yr\left(\frac{M_\bullet}{10^8\msun}\right)^{1.318}\frac{1\msun}{m_\mathrm{eff}}\frac{15}{\log\Lambda},
  \quad r \ll r_\mathrm{dyn}.
\label{eq:t2b}
\end{align}
Note that for the value of $\gamma$ we have chosen the relaxation time
is independent of radius. The Coulomb logarithm is
$\log\Lambda\simeq\log(M_\bullet/m_\mathrm{eff})$ and the effective
stellar mass is $m_\mathrm{eff}=\langle m^2\rangle/\langle m\rangle$
where $\langle\cdot\rangle$ represents a number-weighted average over
the local stellar population. Unfortunately the appropriate value for
$m_\mathrm{eff}$ is quite uncertain. For the solar neighborhood
$m_\mathrm{eff}=0.66\msun$, while for a Salpeter mass function
$m_\mathrm{eff}=0.54 m_{\max}^{0.65}m_{\min}^{0.35}$ where $m_{\max}$
and $m_{\min}$ are the upper and lower cutoffs to the
distribution. For example, when $m_{\max}=100\msun$ and
$m_{\min}=0.1\msun$, $m_\mathrm{eff}=4.8\msun$ (see \citealt{kt11} for
a fuller discussion). For simplicity, in the estimates below we use
$m_\mathrm{eff}=1\msun$, but the actual value of the effective mass is
probably the largest single uncertainty in the estimates of this
subsection.

The resonant relaxation time-scale at semimajor axis $a$ may be written
\citep{ha06,kt11,bof18} 
\begin{align}
t_\mathrm{rr}(a,e)&\simeq 10 \frac{M_\bullet |\dot\omega(a,e)|a^3}{GM_\star(a)
  m_\mathrm{eff}} , \quad r\ll r_\mathrm{dyn}\nonumber \\
&=1.49\times10^{12}\yr \left(\frac{M_\bullet}{10^8\msun}\right)^{1/2} 
\left(\frac{a}{1\pc}\right)^{3/2}\frac{1\msun}{m_\mathrm{eff}}\left|w(e)+
  \frac{3\egr(a)}{1-e^2}\right|.
\label{eq:trr}
\end{align}
The total precession rate can have either sign (retrograde if
self-gravity dominates, prograde if relativistic effects dominate) and
vanishes on the locus where $w(e)<0$ and 
\begin{equation}
r_0(M_\bullet,e)=\frac{0.0378\pc}{[(1-e^2)|w(e)|]^{0.4}}\left(\frac{M_\bullet}{10^8\msun}\right)^{0.727}
\label{eq:r0}
\end{equation}
On this locus the relativistic parameter is
$\egr=\frac{1}{3}(1-e^2)|w(e)|$. 

Stars are lost from the cluster if they pass too close to the central
black hole \citep[][\S 7.5.9]{bt08}. A star crosses the event horizon of a non-rotating black hole if its
pericentre distance $q=a(1-e)<4r_\mathrm{Sch}$, where the
Schwarzschild radius $r_\mathrm{Sch}$ is defined in equation
(\ref{eq:rsch})\footnote{Here the semimajor axis $a$ and eccentricity
  $e$ are determined from the position and velocity at radii much
  larger than $r_\mathrm{Sch}$.}. The star is tidally disrupted if
$q<gR_\star(M_\bullet/m)^{1/3}$ where $R_\star$ is the stellar radius
and $g$ is a factor of order unity. The
fraction of phase space occupied by orbits with pericentre distance
less than $q$ is $2q/a$ if $q\ll a$. Thus the fraction of phase space
on which stars are lost at their next pericentre passage (the `loss
cone') is
\begin{equation}
f_\mathrm{lc}(a)=\max\left[\frac{16GM_\bullet}{c^2a},2g\frac{R_\star}{a}\left(\frac{M_\bullet}{m_\mathrm{eff}}\right)^{1/3}\right].
\label{eq:flc}
\end{equation}
When evaluating this formula, we shall assume $g=1.5$, $R_\star=1\,\mathrm{R}_\odot$ and
$m=1\msun$. 

The results in this subsection are illustrated in Figure
\ref{fig:eight}, as described in the caption and the next subsection.

\subsection{Conditions for a phase transition}

\noindent
We have shown that mono-energetic stellar systems can exhibit lopsided
thermal equilibrium states. Here we use the simplified model of
the preceding subsection to explore whether these equilibria are
likely to be present in black-hole star clusters (and in simulations of
them).

To establish a maximum-entropy equilibrium, the resonant relaxation
time $t_\mathrm{rr}$ must be less than the age of the cluster,
typically $10^{10}\yr$ in an old galaxy (lopsided equilibria are
possible even if this condition is not satisfied, but they reflect the
initial conditions rather than the relaxation process). This region is
shaded in light and dark gray in Figure \ref{fig:eight}. The
derivations in this paper also assume that the resonant relaxation
time is less than the two-body relaxation time $t_{2b}$, a condition
satisfied in the dark gray region; it is likely that our analysis
remains approximately valid even if this condition is violated, since
non-resonant relaxation leads to a steady state that can persist for
many two-body relaxation times. Finally, when the relativistic
parameter $\epsilon_\mathrm{GR}\ga 1$ -- to the left of the white
dashed line, which marks the locus $\egr=0.1666$ -- most of the stars
in the maximum-entropy state have eccentricities near unity (see Fig.\
\ref{fig:five}). In this case our analysis is of limited value since  (i)
thermal equilibrium may not be achieved, since resonant relaxation is
suppressed by rapid precession (the `Schwarzschild barrier'); (ii) the
equilibria may be short-lived because stars on high-eccentricity
orbits are likely to be consumed by the black hole. 

These approximate arguments suggest that the maximum-entropy state is
most likely to be established for black-hole masses $M_\bullet
\la 10^{7.5}\msun$ at radii $\sim 0.001$--$0.1\pc$. 

The maximum-entropy state will be lopsided if the system is
sufficiently `cold', i.e., if the non-Keplerian energy is small enough
(see Figure \ref{fig:four}). In other words, a cluster with an initial
spherically symmetric DF $f(L)$ (eq.\ \ref{eq:mono}) may or may not
suffer a transition to a lopsided state, depending on its energy $E$
as defined by equation (\ref{eq:edef}). For brevity, focus on the case
where relativistic precession is negligible, $\egr=0$. Then the
maximum-entropy state is lopsided for $E<-0.41\,GM_\star^2/a_0$ (Fig.\
\ref{fig:one}), corresponding to spherically symmetric maximum-entropy
systems with rms eccentricity $\langle e^2\rangle^{1/2}< 0.46$ (Fig.\
\ref{fig:two}). We do not understand how black-hole star clusters
form, so we cannot predict whether energies or rms eccentricities in
this range are common. However, the following considerations are
relevant:

\begin{enumerate}

\item The usual default assumption is that the DF in black-hole star
  clusters is ergodic ($\beta=0$), which corresponds to
  $E=-0.3559\, GM_\star^2/a_0$ or
  $\langle e^2\rangle^{1/2}=2^{-1/2}=0.7071$ if $e_{\max}=1$ and the
  system is spherical. For this energy there is no phase
  transition. However, there is no compelling theoretical reason why
  the initial DF of a cluster should be ergodic. 

\item The velocity distribution of the old stars in the black-hole star
  cluster of the Milky Way is close to isotropic \citep{sch09};
  however, the data are reliable only outside a few arcsec (1
  arcsec=$0.04\pc$), well outside the region in which any lopsided
  transition is likely to occur. We are not aware of any direct evidence that
  the old stars near the centre of the Galaxy have a lopsided distribution. 

\item Destruction of stars that pass too close to the black hole tends
  to reduce the cluster energy, since stars on high-eccentricity orbits have
  larger energies in the gravitational potential of the cluster than
  those on low-eccentricity orbits with the same semi-major axis, but
  in the shaded region of Figure \ref{fig:eight} this effect is
  relatively small\footnote{The fractional area of the loss cone in
  phase space is $f_\mathrm{lc}(a)=1-e_{\max}^2$ where $e_{\max}$ is
  the maximum eccentricity if the loss cone is empty (eq.\
  \ref{eq:flc}). Then the dependence of the steady-state DF on
  eccentricity is approximately given by
  $f(e^2)=\log[(1-e^2)/(1-e_{\max}^2)]$ for $e<e_{\max}$ and zero
  otherwise \citep[e.g.,][]{ck78}. For this DF the rms eccentricity is 0.685 for
  $f_\mathrm{lc}=10^{-4}$, 0.677 for $f_\mathrm{lc}=10^{-3}$, and
  0.657 for $f_\mathrm{lc}=10^{-2}$.}.

\item If the stars in the cluster form {\it in situ} then they
  probably form in a disc, although the orientation of the disc and
  the star-formation rate may vary strongly with time. In this case
  the orientations of the stars relax through resonant relaxation much
  faster than the eccentricities (these separate processes are
  sometimes called vector and scalar resonant relaxation,
  respectively; see for example Fig.\ 1 of \citealt{kt11} or Fig.\ 4
  of \citealt{bof18}). In this case the initial state for the
  cluster -- on time-scales longer than the vector resonant relaxation
  time-scale but shorter than the scalar time-scale -- would contain 
  randomly oriented stellar orbits with low eccentricities, and thus
  would be susceptible to the phase transition.

\item An alternative possibility is that black-hole star clusters form
  from the inspiral of globular clusters through dynamical friction
  and their subsequent tidal disruption \citep{tre75,ant12,gne14}. Since dynamical friction
  tends to circularize the globular-cluster orbits, this mechanism would add stars
  to the black-hole star cluster on low-eccentricity orbits. Numerical
  simulations of this process \citep{ant12} show that the resulting
  cluster can have an anisotropy parameter
  $\beta_v=1-\half\sigma_t^2/\sigma_r^2$ (here $\sigma_t$ and $\sigma_r$
  are the velocity dispersions in the radial and tangential
  directions; $\beta_v=0$ for an ergodic system) as
  small as $\sim -0.4$, indicating a significant bias towards
  low-eccentricity orbits which would make the lopsided transition
  more likely. 

\end{enumerate}

Most of our theoretical understanding of the dynamics of black-hole
star clusters is based on analytic arguments or numerical solutions of
the Fokker--Planck equation. These assume spherical symmetry and
therefore do not address the question of whether a lopsided transition
occurs. Direct N-body simulations are much more challenging: only a
few have been carried out, and these remain oversimplified in several
respects. In particular they generally contain too few stars and do
not span the full dynamical range of $\sim 10^6$ between the event
horizon $r_\mathrm{Sch}$ and the dynamical radius $r_\mathrm{dyn}$
(Fig.\ \ref{fig:eight}). As an example we describe the recent
state-of-the-art simulation by \citet{bau18}, which is scaled to the
Milky Way's black-hole star cluster with
$M_\bullet=4\times 10^6\msun$. The simulation contains
$0.95\times 10^6$ stars with a total mass of $4\times 10^7\msun$, so
the mean stellar mass is $\langle m\rangle=42\msun$, which is
unrealistically high. The dynamical radius, where the enclosed stellar
mass equals the black-hole mass, is $r_\mathrm{dyn}\simeq 1.2\pc$ at
the start of the simulation, growing slowly to $1.5\pc$ over
$5.5\mbox{\,Gyr}$. From Figure \ref{fig:eight}, we expect the resonant
relaxation time-scale to be shorter than the two-body time-scale (dark
gray band) at radii less than about $0.04\pc$, and this is where any
lopsided transition is expected to occur\footnote{This simulation does
  not include relativistic effects and so lopsided equilibria can
  persist at smaller radii than indicated in Figure
  \ref{fig:eight}.}. However, within this radius the \citet{bau18}
simulation has only $\sim 10-20$ stars. This unrealistically small
number arises for two reasons: (i) a large assumed mean stellar mass,
which reduces the number of stars at each radius and enhances the
relaxation rate; (ii) an artificially large Schwarzschild radius for
the black hole, $10^3$ times its actual value or $0.0004\pc$, which
enhances the consumption rate of the black hole and leads to a
shallower central density cusp than would otherwise be present. Thus
it is unlikely that a detectable lopsided region would arise in this
simulation, even if one were to be expected in a fully realistic
N-body simulation of the same cluster.

\section{Summary}

\label{sec:summary}

\noindent
We have explored the equilibria of stellar systems orbiting in the
gravitational field of a central massive object, typically a black
hole. In particular we have focused on the thermal equilibria over
time-scales long compared to the resonant relaxation time and short
compared to the two-body relaxation time. These equilibria maximize
the entropy subject to the usual constraints that the mass, energy,
and angular momentum are conserved, and the additional constraint that
the semimajor axes of the stars are conserved.

We have shown that this system exhibits a phase transition from a
disordered high-temperature equilibrium state to an ordered
low-temperature state. The disordered state is spherically symmetric,
while in the ordered state the stellar orbits have higher
eccentricities and nearly aligned apsides. The `temperature' of the
system is a measure of the non-Keplerian component of its total
energy, which arises from the self-gravity of the system and any
relativistic corrections to the Keplerian Hamiltonian. In the absence
of relativistic precession, the lopsided states correspond to systems
with small self-gravitational energy, $E<-0.41\,GM_\star^2/a_0$, which
arise from initially spherical states with relatively small rms
eccentricity, $\langle e^2\rangle^{1/2}< 0.46$.

The existence of lopsided equilibria in stellar systems dominated by a
central black hole is not too surprising: the black-hole star cluster in
M31 is lopsided \citep{tre95,pt03,bm13}, both analytic and N-body
models of nearly Keplerian discs exhibit lopsided secular
instabilities \citep{js01,touma02,ss10,ts12}, and it is
straightforward to construct equilibrium models of collisionless or
fluid eccentric discs \citep{statler01,ob14,dr18,ldl18}. In contrast,
the lopsided equilibria described here bifurcate from spherical
systems rather than axisymmetric discs, and have no precession or
rotation. 

For simplicity we have specialized to the case of a single stellar
mass, and to a mono-energetic system -- by which we mean that all
stars have the same semimajor axis -- but many of our conclusions also
hold for systems with a more realistic distribution of masses and
semimajor axes \citep{ttk19,tre19}.

We have only looked carefully at axisymmetric lopsided systems, that is,
systems that are axisymmetric around the $z$-axis but asymmetric with
respect to the $z=0$ plane. Non-axisymmetric maximum-entropy systems
may also be present but we have not found any. 

The lopsided equilibria persist when relativistic precession is
present, but when the relativistic parameter $\egr\ga 1$ (eq.\
\ref{eq:epsgr}) these have rms eccentricity near unity and thus our
models are unrealistic, since resonant relaxation is suppressed by the
Schwarzschild barrier and we have not accounted for the loss of stars
on high-eccentricity orbits that pass close to the black hole. The
high rms eccentricities found in lopsided states should enhance the
rate of tidal disruption events and extreme mass-ratio inspirals,
which may be detected through optical, ultraviolet, or X-ray transient
searches or gravitational-wave observatories.

The phase transition described here is driven by scalar resonant
relaxation, in which the eccentricities and orientations of the orbits
relax while their semimajor axes remain fixed. In vector resonant
relaxation, the eccentricities and semimajor axes remain fixed and
only the orientations of the orbital planes relax. Vector resonant
relaxation can also drive phase transitions \citep{roupas17}.

The lopsided equilibria we have found are both dynamical (solutions of the
collisionless Boltzmann equation\footnote{Strictly, we have only
  established that equilibrium exists in the orbit-averaged sense,
  i.e., that the system satisfies the collisionless Boltzmann equation
  after averaging over the characteristic dynamical time $\Omega^{-1}$
  (eq.\ \ref{eq:omdef}), but this approximation should be benign so
  long as $M_\star \ll M_\bullet$.}) and thermal (global maxima of the
entropy, subject to the constraint that the stellar semimajor axes are
fixed in resonant relaxation\footnote{Thus they are thermal equilibria
  on time-scales long compared to the resonant relaxation time-scale but
  short compared to the two-body relaxation time-scale.}). Thus the
lopsided equilibria are {\it possible} even if resonant
relaxation is not complete, but {\it required} if it is.

Important next steps are to establish that lopsided equilibria are
found in simulations of secular dynamics (Touma \& Kazandjian, in
preparation) and in direct N-body simulations of star clusters
containing central black holes.  Realistic simulations of black-hole
star clusters are challenging but the results of this paper can be
used to guide the design of simpler N-body simulations that should
still exhibit the relevant behaviour.

\section*{Acknowledgements}

\noindent
We thank Ben Bar-Or and Jean-Baptiste Fouvry for comments and
discussions that improved our understanding and presentation. This
research emerged from discussions with Jihad Touma about instabilities
in simulations of mono-energetic black-hole star clusters, and would not have been
possible without his insights and encouragement.

\appendix

\section{Linear stability of spherical maximum-entropy systems}

\label{sec:stable}

\noindent
To keep the derivations in this section as general as possible, we do
not use the assumptions of a mono-energetic system and of axisymmetry
that we introduced at the start of \S\ref{sec:mono}. 

\subsection{Thermodynamic stability}

\noindent
In a spherically symmetric system the DF can depend only on the
integrals of motion $\Lambda$ and $L$. Thus the DF of a perturbed
spherical system can be written
\begin{equation}
F(\Lambda,\bfL,\bftheta)=F_0(\Lambda,L) + \Delta
F(\Lambda,\bfL,\bftheta)
\end{equation}
where $|\Delta F|/F_0 \ll1$. Expanding equation (\ref{eq:entropy}) to
$\mbox{O}(\Delta F)^2$ we find that the perturbation to the entropy is
\begin{align}
\Delta S&=-2\pi\int \rd\Lambda \rd\bfL \rd\bftheta\big[
F(\Lambda,\bfL,\bftheta)\log
F(\Lambda,\bfL,\bftheta)-F_0(\Lambda,L)\log
F_0(\Lambda,L)\big]\nonumber \\
&=-2\pi \int \rd\Lambda \rd\bfL \rd\bftheta\left[\Delta F +
\Delta F\log F_0 +\frac{\Delta F^2}{2F_0}\right]
+\mbox{O}(\Delta F)^3.
\label{eq:ds}
\end{align}

Similarly, from equation (\ref{eq:edef}) the perturbed energy is
\begin{align}
\Delta E=2\pi\int \rd\Lambda \rd\bfL \rd\bftheta\, \Delta F(\Lambda,\bfL,\bftheta)
H_0(\Lambda,L) + 2\pi^2\int \rd\Lambda \rd\bfL \rd\bftheta \rd\Lambda' \rd\bfL'
\rd\bftheta'\, \Delta F(\Lambda,\bfL,\bftheta) \mathbfss{K}(\Lambda,\Lambda',\bfL,\bfL',\bftheta,\bftheta') \Delta
  F(\Lambda',\bfL',\bftheta') +\mbox{O}(\Delta F)^3,
\label{eq:edefpp}
\end{align}
in which we have defined the unperturbed Hamiltonian (cf.\ eq.\ \ref{eq:hdef})
\begin{equation}
H_0(\Lambda,L)=H_\mathrm{GR}(\Lambda,L) +2\pi\int \rd\Lambda' \rd\bfL'
\rd\bftheta'\, \mathbfss{K}(\Lambda,\Lambda',\bfL,\bfL',\bftheta,\bftheta')
  F_0(\Lambda',L').
\end{equation}
The perturbation to the mass at a given semimajor axis (eq.\
\ref{eq:rhodef}) is
\begin{equation}
\Delta \rho(\Lambda)=2\pi\int \rd\bfL \rd\bftheta\, \Delta
F(\Lambda,\bfL,\bftheta).
\end{equation}

Since the unperturbed DF is an equilibrium, it must be an extremum of the entropy at
fixed energy ($\Delta E=0$), and since resonant relaxation conserves
semimajor axis we also require $\Delta\rho(\Lambda)=0$. These
requirements are satisfied if $\Delta S
-\beta\Delta E -\gamma(\Lambda) \Delta \rho(\Lambda)=\mbox{O}(\Delta F)^2$ for all
variations $\Delta F(\Lambda,\bfL,\bftheta)$, where $\beta$ and
$\gamma(\Lambda)$ are Lagrange multipliers. This condition implies that
$F_0(\Lambda,L)=\exp[-\beta H_0(\Lambda,L)-\gamma(\Lambda)-1]$ (cf.\ eq.\ \ref{eq:betadef}). Substituting this
result into equation (\ref{eq:ds}) yields
\begin{equation}
\Delta S=2\pi \int \rd\Lambda \rd\bfL \rd\bftheta\left[\gamma(\Lambda)\Delta F + \beta H_0\Delta F -\frac{\Delta F^2}{2F_0}\right]
+\mbox{O}(\Delta F)^3.
\label{eq:dss}
\end{equation}
Using the conditions $\Delta \rho(\Lambda)=0$ and $\Delta E=0$ to 
eliminate the terms linear in $\Delta F$, we find 
\begin{align}
\Delta S=-2\pi^2\beta \int \rd\Lambda \rd\bfL \rd\bftheta \rd\Lambda' \rd\bfL'
\rd\bftheta'\, \Delta F(\Lambda,\bfL,\bftheta)
\mathbfss{K}(\Lambda,\Lambda',\bfL,\bfL',\bftheta,\bftheta') \Delta
  F(\Lambda',\bfL',\bftheta') -\pi \int \rd\Lambda \rd\bfL \rd\bftheta\,\frac{\Delta F^2}{F_0}
+\mbox{O}(\Delta F)^3.
\label{eq:dsss}
\end{align}
The system is thermodynamically stable if the entropy is a
local maximum, which requires that $\Delta S\le 0$ for all variations
$\Delta F$ that conserve $E$ and $\rho(\Lambda)$. 

We can write the dependence of $\Delta F$ on the orientation angles
$\bftheta=(\omega,\Omega)$ and $I=\cos^{-1}L_z/L$ as an expansion in
Wigner d-matrices (\ref{eq:wigner}), since these provide a complete representation of
the rotation group SO(3):
\begin{equation} 
\Delta F(\Lambda,\bfL,\bftheta)=\sum_{l=0}^\infty \sum_{m=-l}^l \sum_{n=-l}^l
B^{\,l}_{nm}(\Lambda,L) d^{\,l}_{nm}(I)\re^{\ri n\omega +\ri m\Omega}.  
\label{eq:fexpand}
\end{equation} 
Here $\{B^{\,l}_{nm}(\Lambda,L)\}$ are the undetermined functions
that specify $\Delta F$. Since $\Delta F$ is real, the relation
(\ref{eq:symm}) implies that
\begin{equation}
B^{\,l\,\,*}_{-n-m}(\Lambda,L)=(-1)^{m-n}B^{\,l}_{nm}(\Lambda,L).
\label{eq:symm1}
\end{equation}
 Using equation (\ref{eq:kkred}) and the
orthogonality relation (\ref{eq:ortho}) we find
\begin{align}
\Delta S &=-4\pi^3\sum_{l=0}^\infty\sum_{m=-l}^l\sum_{n=-l}^l \int  \frac{\rd\Lambda
 \rd L^2}{2l+1}\,\frac{|B^{\,l}_{nm}(\Lambda,L)|^2}{F_0(\Lambda,L)}
 -(2\pi)^7G\beta\sum_{l=0}^\infty\sum_{m=-l}^l\sum_{n,n'=-l}^l\!\!\frac{\ri^{n-n'}y_{ln}y_{ln'}}{(2l+1)^3}
  \nonumber \\
&\qquad \times \int
  \rd\Lambda \rd L^2 \rd\Lambda' d{L'}^2\,{B^{\,l}}_{nm}^*(\Lambda,L)B^{\,l}_{n'm}(\Lambda',L')
  \mathbfss{Q}^{\,l}_{nn'}(\Lambda,L,\Lambda',L'). 
\end{align}
Thus the system is thermodynamically stable if 
\begin{align}
 &\frac{2^5\pi^4G\beta}{\sum_{l=0}^\infty\sum_{m=-l}^l\sum_{n=-l}^l (2l+1)^{-1}\int  \rd\Lambda
 \rd L^2\,|B^{\,l}_{nm}(\Lambda,L)|^2/F_0(\Lambda,L)}\sum_{l=0}^\infty\sum_{m=-l}^l\sum_{n,n'=-l}^l
\nonumber \\
&\quad\times \frac{\ri^{n-n'}y_{ln}y_{ln'}}{(2l+1)^3}\!\int
  \rd\Lambda \rd L^2 \rd\Lambda' d{L'}^2{B^{\,l}}_{nm}^*(\Lambda,L)B^{\,l}_{n'm}(\Lambda',L')
  \mathbfss{Q}^{\,l}_{nn'}(\Lambda,L,\Lambda',L')  > -1
\label{eq:stabt}
\end{align}
for all trial functions $B^{\,l}_{nm}(\Lambda,L)$. Note that (i) this
is a sufficient condition for stability; a necessary condition is that
the inequality is satisfied for all trial functions for which
$\Delta\rho(\Lambda)=0$, which in turn requires
$\int \rd L^2\, B^0_{00}(\Lambda,L)=0$; (ii)
$y_{ln}=Y_{ln}(\half\pi,0)=0$ unless $l-n$ is even, so only terms with
even values of $l-n$ and $l-n'$ need to be considered; (iii) the
criterion is independent of $m$ except for the trial function
$B^{\,l}_{nm}$ (as it must be, since the equilibrium system is
spherically symmetric), so the sum over $m$ in equation
(\ref{eq:stabt}) can be dropped; (iv) the stability criterion is
satisfied for an arbitrary set of trial functions $B^{\,l}_{nm}$ if
and only if it is satisfied for a restricted set of functions in which
$B^{\,l}_{nm}$ is non-zero for only one value of $l$, so the sum over
$l$ can be dropped if the inequality is satisfied for every 
$l$; (v) the relativistic Hamiltonian enters the stability criterion
only through its effect on the equilibrium DF $F_0(\Lambda,L)$.
Moreover, (vi) the sums can be shortened to sums over non-negative $n$
by observing that $\mathbfss{Q}^{\,l}_{nn'}$ is even in both $n$ and
$n'$ and that $y_{l,-n}=(-1)^ny_{ln}$, and by writing
$B^{\,l}_{\pm n,m}\equiv \half \ri^{|m|}(S_{n}\pm A_{n})$ with $n\ge0$; together
with equation (\ref{eq:symm1}) this implies that $S_n$ is real if $l$
and $n$ are even and imaginary if $l$ and $n$ are odd, with the
opposite true for $A_n$. Thus the thermodynamic stability criterion is
simplified to
\begin{align}
 &\frac{2^6\pi^4G\beta}{\sum_{n=0}^l (1+\delta_{n0})^{-1}\int  \rd\Lambda
 \rd L^2\,\big(|S_{n}(\Lambda,L)|^2+|A_{n}(\Lambda,L)|^2\big)/F_0(\Lambda,L)}\nonumber \\
&\times 
\sum_{n,n'=0}^l \ri^{n-n'}c_{ln}c_{ln'}\int
  \rd\Lambda \rd L^2 \rd\Lambda' d{L'}^2\,S_{n}^*(\Lambda,L)S_{n'}(\Lambda',L')
  \mathbfss{Q}^{\,l}_{nn'}(\Lambda,L,\Lambda',L')  > -1, \quad \forall \ell,
\label{eq:stab5}
\end{align}
where $c_{ln}$ is defined in equation (\ref{eq:cdef}). 
This formula shows that the most unstable perturbations
are those with $A_n(\Lambda,L)=0$ so we assume this from now on. 

For numerical work we discretize these integrals by covering 
$(\Lambda,L^2)$ space with bins centred on
$(\Lambda_\alpha,L^2_\alpha)$ having area $\Delta
\Lambda_\alpha\Delta L^2_\alpha$, $\alpha=1,2,\ldots$. We define
\begin{align}
s_{\alpha n}&\equiv
\ri^{-n}S_n(\Lambda_\alpha,L_\alpha)\left[\frac{\Delta\Lambda_\alpha
 \Delta L^2_\alpha}{(1+\delta_{n0})F_0(\Lambda_\alpha,L_\alpha)}\right]^{1/2}\nonumber\\
\mathbfss{R}^{\,l}_{\alpha n,\alpha'n'}&\equiv 2^6\pi^4G\beta c_{ln}c_{ln'}\left[\Delta
 \Lambda_\alpha\Delta L^2_\alpha F_0(\Lambda_\alpha,L_\alpha) \Delta
  \Lambda_{\alpha'}\Delta L^2_{\alpha'}F_0(\Lambda_{\alpha'},L_{\alpha'})
\right]^{1/2}[(1+\delta_{n0})(1+\delta_{n'0})]^{1/2}\,\mathbfss{Q}^{\,l}_{nn'}(\Lambda_\alpha,L_\alpha,\Lambda_{\alpha'},L_{\alpha'}).
\label{eq:discrete}
\end{align}
The stability criterion (\ref{eq:stab5}) becomes
\begin{equation}
\frac{\sum_{nn'\ge0}\sum_{\alpha\alpha'} s^*_{\alpha
  n}\mathbfss{R}^{\,l}_{\alpha n,\alpha'n'}
s_{\alpha'n'}}{\sum_{nn'\ge0}\sum_\alpha |s_{\alpha n}|^2} > -1.
\label{eq:ineq2}
\end{equation}
Now $\mathbfss{R}^{\,l}_{\alpha n,\alpha' n'}$ is a real symmetric 
matrix with multi-index $(\alpha,n)$ so its eigenvalues are
real. Since (\ref{eq:ineq2}) is the
Rayleigh quotient of $\mathbfss{R}^{\,l}$, its minimum is the
smallest eigenvalue of $\mathbfss{R}^{\,l}$, say, $\lambda^{\,l}_\mathrm{min}$, and
the system is thermodynamically stable if 
$\lambda^{\,l}_\mathrm{min}>-1$.  This sufficient condition is also
necessary if the perturbation conserves mass, $\int \rd\Lambda \rd\bfL
\rd\bftheta\,\Delta F =0$. This requirement is automatically satisfied
if $l>0$. 

For the mono-energetic systems explored in this paper, the spherically
symmetric equilibrium DF has the form
$F_0(\Lambda,L)=\delta(\Lambda-\Lambda_0)f_0(L)$ (cf.\ eq.\
\ref{eq:mono}). To evaluate the matrix $\mathbfss{R}^{\,l}$ we need the equilibrium
spherical DF $f_0(L)$. We find this by evaluating $f$ using
equation (\ref{eq:betadef}) for an assumed Hamiltonian $H$, then
evaluating $H$ using equation (\ref{eq:hdef}), and iterating to
convergence. 

\subsection{Dynamical stability}

\noindent
The linearized collisionless Boltzmann equation that describes the
evolution of small perturbations to an equilibrium DF is
\begin{equation}
\frac{\p\Delta F}{\p t} +\frac{\p H_0}{\p \bfJ}\cdot\frac{\p \Delta
  F}{\p\bfw}-\frac{\p \Delta H}{\p \bfw}\cdot\frac{\p F_0}{\p\bfJ}=0.
\label{eq:cbe}
\end{equation}
Here $\bfJ=(\Lambda,L,L_z)$ is a vector of the three actions, 
$\bfw=(\ell,\omega,\Omega)$ is a vector of the angles, and $\Delta H$
is the perturbed Hamiltonian. Since we are working in the secular
approximation, $\Delta F$ and $\Delta H$ are independent of the mean
longitude $\ell$. Since the equilibrium DF is spherically symmetric,
$F_0$ and $H_0$ are independent of $L_z$. Moreover $\Delta F$ can be
represented as a linear combination of functions with time dependence
$\exp(-\ri\omega t)$. Thus equation (\ref{eq:cbe})
simplifies to
\begin{equation}
-\ri\omega\Delta F +\frac{\p H_0}{\p L}\frac{\p \Delta
  F}{\p\omega}-\frac{\p \Delta H}{\p \omega}\frac{\p F_0}{\p L}=0.
\label{eq:cbe1}
\end{equation}

The perturbed Hamiltonian is (cf.\ eq.\ \ref{eq:hdef})
\begin{equation}
\Delta H(\Lambda,\bfL,\bftheta)=2\pi \int
\rd\Lambda'\rd\bfL'\rd\bftheta'\,\mathbfss{K}(\Lambda,\Lambda',\bfL,\bfL',\bftheta,\bftheta')\Delta
F(\Lambda',\bfL',\bftheta').
\end{equation}
Using equations (\ref{eq:ortho}) and (\ref{eq:kkred}) as well as the
expansion of the perturbed DF (\ref{eq:fexpand}), we find
\begin{align}
\Delta H(\Lambda,\bfL,\bftheta)&=2^5\pi^4G
\sum_{l=0}^\infty\sum_{m=-l}^l\sum_{nn'=-l}^l\frac{\ri^{n-n'}y_{ln}y_{ln'}}{(2l+1)^2}d^{\,l}_{nm}(I)\re^{\ri
 m\Omega+in\omega}
\int \rd\Lambda'd{L'}^2\,\mathbfss{Q}^{\,l}_{nn'}(\Lambda,L,\Lambda',L') B^{\,l}_{nm}(\Lambda',L').
\end{align}
Because of the orthogonality relation (\ref{eq:ortho}) we can examine
terms of a single $l$ and $m$ in equation (\ref{eq:cbe1}). Thus we can drop the
indices $l$ and $m$ on $B^{\,l}_{nm}$ and the linearized collisionless
Boltzmann equation now reads 
\begin{align}
-\omega B_n + n\frac{\p H_0}{\p L}B_n - 2^5\pi^4nG\frac{\p F_0}{\p L}\sum_{n'=-l}^l
\frac{\ri^{n-n'}y_{ln}y_{ln'}}{(2l+1)^2} \int \rd\Lambda'
d{L'}^2\mathbfss{Q}_{nn'}^{\,l}(\Lambda,L,\Lambda',L')B_{n'}(\Lambda',L')=0.
\end{align}

As in the preceding subsection, we can split this equation into components
that are even and odd in $n$ by writing $B_{\pm n}=\half \ri^{|m|}(S_n\pm A_n)$
with $n\ge0$. Since $\mathbfss{Q}^{\,l}_{nn'}$, $y_{ln}\ri^n$, and $y_{ln}\ri^{-n}$ are
all even in $n$, for $n\ge0$ we have 
\begin{align}
\omega S_n&=n\frac{\p H_0}{\p L}A_n, \nonumber \\
\omega A_n&=n\frac{\p H_0}{\p L}S_n -2^6\pi^4 nG\frac{\p F_0}{\p
  L} \sum_{n'\ge0}\frac{\ri^{n-n'}y_{ln}y_{ln'}}{(2l+1)^2(1+\delta_{n'0})}\int  \rd\Lambda'
d{L'}^2\mathbfss{Q}_{nn'}^{\,l}(\Lambda,L,\Lambda',L')S_{n'}(\Lambda',L').
\end{align}
Eliminating $A_n$,
\begin{align}
n^2\left(\frac{\p H_0}{\p L}\right)^2S_n -2^6\pi^4 n^2G\frac{\p H_0}{\p
  L}\frac{\p F_0}{\p
  L}\sum_{n'\ge0}\frac{\ri^{n-n'}y_{ln}y_{ln'}}{(2l+1)^2(1+\delta_{n'0})}  
\int  \rd\Lambda' d{L'}^2\mathbfss{Q}_{nn'}^{\,l}(\Lambda,L,\Lambda',L')S_{n'}(\Lambda',L')=\omega^2
  S_n. 
\label{eq:stab}
\end{align}
If the unperturbed DF is in thermal equilibrium then at a given
semimajor axis $F_0\propto\exp(-\beta H_0)$, so $\p F_0/\p L=-\beta
F_0\, \p H_0/\p L$ and 
\begin{align}
n^2\left(\frac{\p H_0}{\p L}\right)^2S_n +&2^6\pi^4 n^2G\beta\left(\frac{\p H_0}{\p
  L}\right)^2
  F_0\sum_{n'\ge0}\frac{\ri^{n-n'}y_{ln}y_{ln'}}{(2l+1)^2(1+\delta_{n'0})}
\int
  \rd\Lambda'd{L'}^2\mathbfss{Q}_{nn'}^{\,l}(\Lambda,L,\Lambda',L')S_{n'}(\Lambda',L')=\omega^2
  S_n. 
\end{align}
We now discretize this integral equation as in equations
(\ref{eq:discrete}):
\begin{equation}
\sum_{\alpha'n'}\mathbfss{V}^{\,l}_{\alpha n,\alpha'n'}s_{\alpha'n'}=\omega^2
s_{\alpha n}
\label{eq:vdefa}
\end{equation}
where
\begin{equation}
\mathbfss{V}^{\,l}_{\alpha n,\alpha'n'}\equiv n^2\left(\frac{\p H_0}{\p
  L}\right)^2_{\alpha}\left[\delta_{nn'}\delta_{\alpha\alpha'}+\mathbfss{R}^{\,l}_{\alpha
  n,\alpha'n'}\right]
\label{eq:vdef}
\end{equation}
Thus $\omega^2$ is an eigenvalue of the matrix $\mathbfss{V}^{\,l}$, which
is closely related to the matrix $\mathbfss{R}^{\,l}$ whose eigenvalues
determine thermodynamic stability. The matrix $\mathbfss{V}^{\,l}$ is real
and although it is non-symmetric its eigenvalues are all
real (see below for proof). Thus the system is dynamically unstable if and
only if the minimum eigenvalue is negative, $\omega^2<0$.

\subsection{Proof that eigenvalues of $\mathbfss{V}^l$ are real}

\noindent
Rewrite (\ref{eq:vdefa}) and (\ref{eq:vdef}) as
$\mathbfss{V}\bfs =\omega^2\bfs$ where
$\mathbfss{V}=\mathbfss{D} +\mathbfss{D}\mathbfss{R}$, with $\mathbfss{D}$ a
diagonal matrix having entries $n^2(\p H_0/\p L)_\alpha^2$.  Rearrange
the labeling of the rows and columns in $\mathbfss{D}$ and $\mathbfss{R}$
so the first $N$ diagonal elements of $\mathbfss{D}$ are non-zero and the last $M$
elements are zero. Then we can write $\mathbfss{D}$ and $\mathbfss{R}$ in
block form as
\begin{equation}
\mathbfss{D}=\left[\begin{array}{cc} \mathbfss{D}_1 & {\bf 0} \\ {\bf 0} &
 {\bf 0} \end{array}\right] \quad
\mathbfss{R} = \left[\begin{array}{cc} \mathbfss{R}_a & \mathbfss{R}_b \\
 \mathbfss{R}^t_b & \mathbfss{R}_c \end{array}\right];
\end{equation}
here $\mathbfss{D}_1$ is a positive-definite $N\times N$ diagonal
matrix, $\mathbfss{R}_a$ is a symmetric real $N\times N$ matrix, 
$\mathbfss{R}_c$ is a symmetric real $M\times M$ matrix, 
$\mathbfss{R}_b$ is a real $M\times N$ matrix, and $\mathbfss{R}^t_b$ is
its transpose. Writing $\bfs^t=[\bfx^t \ \bfy^t]$ where $\bfx$ and
$\bfy$ are $1\times N$ and $1\times M$ column matrices, the
eigenvalue equation becomes 
\begin{equation}
\mathbfss{D}_1\mathbfss{R}_a\bfx +\mathbfss{D}_1\bfx +\mathbfss{D}_1\mathbfss{R}_b\bfy =
 \omega^2\bfx,  \quad {\bf 0}=\omega^2\bfy.
\end{equation}
Therefore either $\omega^2=0$ or $\bfy=0$. In the latter case, the
  eigenvalue equation becomes
\begin{equation}
\mathbfss{D}_1\mathbfss{R}_a\bfx +\mathbfss{D}_1\bfx =\omega^2\bfx.
\end{equation}
Since $\mathbfss{D}_1$ is diagonal with positive-definite diagonal
elements, we can define a $1\times N$ column matrix $\bfz$ whose elements are
$z_n=D_{1,nn}^{-1/2}x_n$ and a symmetric, real $N\times N$ matrix $\mathbfss{W}$ by
$\mathbfss{W}_{jk}=D_{1,jj}^{1/2} \mathbfss{R}_{a,jk} D_{1,kk}^{1/2}$. Then 
\begin{equation}
(\mathbfss{W}+\mathbfss{D}_1)\bfz=\omega^2\bfz.
\end{equation}
Since the matrix on the left side is symmetric, its eigenvalues
$\omega^2$ must be real.

\label{lastpage}


\begin{thebibliography}{99}

\bibitem[Alexander(2017)]{alex17} Alexander T., 2017, \araa, 55, 17 

\bibitem[Antonini et al.(2012)]{ant12} Antonini F.,
  Capuzzo--Dolcetta R., Mastrobuono--Battisti A., Merritt D., 2012,
  \apj, 750, 111  

\bibitem[Bar-Or \& Alexander(2016)]{boa16} Bar-Or B., Alexander T., 2016, \apj, 820, 129 

\bibitem[Bar-Or \& Fouvry(2018)]{bof18} Bar-Or B., Fouvry J.-B., 2018, \apjl, 860, L23 

\bibitem[Baumgardt et al.(2018)]{bau18} Baumgardt H., Amaro-Seoane P., Sch{\"o}del, R., 2018, \aap, 609, A28 

\bibitem[Binney \& Tremaine(2008)]{bt08} Binney J., Tremaine S.,\
  2008, Galactic Dynamics, 2nd ed., Princeton Univ.\ Press,
  Princeton, NJ

\bibitem[Brown \& Magorrian(2013)]{bm13} Brown C.K., Magorrian J.,\ 2013, \mnras, 431, 80 

\bibitem[Cohn \& Kulsrud(1978)]{ck78} Cohn H., Kulsrud R.M.,\
  1978, \apj, 226, 1087 

\bibitem[Davydenkova \& Rafikov(2018)]{dr18} Davydenkova I.,
  Rafikov R.R., 2018, \apj, 864, 74 

\bibitem[Gnedin et al.(2014)]{gne14} Gnedin O.Y., Ostriker J.P., Tremaine S., 2014, \apj, 785, 71 

\bibitem[Hopman \& Alexander(2006)]{ha06} Hopman C., Alexander
  T., 2006, \apj, 645, 1152 

\bibitem[Jacobs \& Sellwood(2001)]{js01} Jacobs V., Sellwood J.A., 2001, \apjl, 555, L25 

\bibitem[Kocsis \& Tremaine(2011)]{kt11} Kocsis B., Tremaine S., 2011, \mnras, 412, 187 

\bibitem[Kormendy \& Ho(2013)]{kh13} Kormendy J., Ho L.C., 2013,  \araa, 51, 511 

\bibitem[Lee et al.(2018)]{ldl18} Lee W.-K., Dempsey A.M., Lithwick Y., 2018, arXiv:1811.11758 

\bibitem[Merritt et al.(2011)]{mer11} Merritt D., Alexander T., Mikkola S., Will C.M., 2011, \prd, 84, 044024 

\bibitem[Ogilvie \& Barker(2014)]{ob14} Ogilvie G.I., Barker A.J., 2014, \mnras, 445, 2621 

\bibitem[Peiris \& Tremaine(2003)]{pt03} Peiris H.V., Tremaine S., 2003, \apj, 599, 237 

\bibitem[Polyachenko et al.(2007)]{pps07} Polyachenko V.L., Polyachenko E.V., Shukhman I.G., 2007, Soviet Journal of Experimental and Theoretical Physics, 104, 396 

 \bibitem[Roupas et al.(2017)]{roupas17} Roupas Z., Kocsis B., Tremaine S.\ 2017, \apj, 842, 90

\bibitem[Sch{\"o}del et al.(2009)]{sch09} Sch{\"o}del R., Merritt D., Eckart A., 2009, \aap, 502, 91 

\bibitem[Sridhar \& Saini(2010)]{ss10} Sridhar S., Saini T.D., 2010, \mnras, 404, 527 

\bibitem[Statler(2001)]{statler01} Statler T.S., 2001, \aj, 122, 2257 

\bibitem[Touma(2002)]{touma02} Touma J.R., 2002, \mnras, 333, 583 

\bibitem[Touma \& Sridhar(2012)]{ts12} Touma J.R., Sridhar S., 2012,
  \mnras, 423, 2083

 \bibitem[Touma, Tremaine \& Kazandjian(2019)]{ttk19} Touma J., Tremaine S., Kazandjian M.,\ 2019, \prl, 123, 021103

\bibitem[Tremaine(1995)]{tre95} Tremaine S.,\ 1995, \aj, 110, 628 

\bibitem[Tremaine(2005)]{tre05} Tremaine S.,\ 2005, \apj, 625, 143

 \bibitem[Tremaine(2019)]{tre19} Tremaine S.,\ 2019, MNRAS, in press

\bibitem[Tremaine et al.(1975)]{tre75} Tremaine S.D., Ostriker J.P., Spitzer L., Jr., 1975, \apj, 196, 407 

\end{thebibliography}
\end{document}